\documentclass[fleqn,usenatbib]{mnras}

\usepackage{newtxtext,newtxmath}

\usepackage[T1]{fontenc}
\usepackage{float}

\DeclareRobustCommand{\VAN}[3]{#2}
\let\VANthebibliography\thebibliography
\def\thebibliography{\DeclareRobustCommand{\VAN}[3]{##3}\VANthebibliography}

\usepackage{graphicx}	
\usepackage{amsmath}	

\usepackage{amssymb}	

\title[The difficult path to coalescence of massive black holes] {The difficult path to coalescence: massive black hole dynamics in merging low mass dark matter haloes and galaxies}

\author[C. Partmann et al.]{Christian Partmann$^{1}$\thanks{E-mail: partmann@mpa-garching.mpg.de},
Thorsten Naab$^{1}$,
Antti Rantala$^{1}$,
Anna Genina$^{1}$,
Matias Mannerkoski$^{2}$, \newauthor and  Peter H. Johansson$^2$
\\
$^{1}$Max Planck Institute for Astrophysik, Karl-Schwarzschild-Str. 1, 85748, Garching\\
$^{2}$Department of Physics, University of Helsinki, Gustaf H\"allstr\"omin katu 2, FI-00014 Helsinki, Finland\\
}

\date{Accepted XXX. Received YYY; in original form ZZZ}

\pubyear{2022}

\begin{document}
\label{firstpage}
\pagerange{\pageref{firstpage}--\pageref{lastpage}}
\maketitle

\begin{abstract}
We present a high resolution numerical study of the sinking and merging of massive black holes (MBHs) with masses in the range of $10^3 - 10^7 \, \mathrm{M}_\odot$ in multiple minor mergers of low mass dark matter halos without and with galaxies ($4\times 10^8 \, \mathrm{M}_\odot \lesssim \mathrm{M}_{\mathrm{halo}} \lesssim 2\times 10^{10} \, \mathrm{M}_\odot)$. The \textsc{Ketju} simulation code, a combination of the \textsc{Gadget} tree solver with accurate regularised integration, uses unsoftened forces  between the star/dark matter components and the MBHs for an accurate treatment of dynamical friction and scattering of dark matter/stars by MBH binaries or multiples. Post-Newtonian corrections up to order 3.5 for MBH interactions allow for coalescence by gravitational wave emission and gravitational recoil kicks. Low mass MBHs ($\lesssim 10^5 \, \mathrm{M}_\odot$) hardly sink to the centre or merge. Sinking MBHs have various complex evolution paths - binaries, triplets, free-floating MBHs, and dynamically or recoil ejected MBHs. Collisional interactions with dark matter alone can drive MBHs to coalescence. The highest mass MBHs of $\gtrsim 10^6 M_\odot$ mostly sink to the centre and trigger the scouring of dark matter and stellar cores. The scouring can transform a centrally baryon dominated system to a dark matter dominated system. Our idealized high-resolution study highlights the difficulty to bring in and keep low mass MBHs in the centres of low mass halos/galaxies – a remaining challenge for merger assisted MBH seed growth mechanisms. 
\end{abstract}

\begin{keywords}
galaxies - black hole physics - black hole mergers - dark matter
\end{keywords}



\section{Introduction}
The formation and growth processes of massive black holes (MBH), exceeding the masses of stellar mass black holes, is still unknown. High-redshift observations have established that supermassive black holes (SMBH is the usual term for black holes with masses $M_\bullet \gtrsim 10^6 \mathrm{M}_\odot$) already existed during the first billion years of the Universe \citep[e.g.][]{Fan_2001, 2003AJ....125.1649F, Lawrence_2007, Willott_2007, 2012AJ....143..142M, Wu_2015, Ba_ados_2017, 2011Natur.474..616M, 2020ApJ...897L..14Y, Wang_2021}. The detection of SMBHs out to redshift $z \sim 11$ \citep{maiolino2023small} poses the question of how they could form and grow so quickly \citep{Inayoshi_2020, bennett2023growth, Costa:2023mgw}. While these observations typically represent the most massive and luminous part of the population ($M_\bullet \sim 10^9 \, \mathrm{M}_\odot$), a larger population of lower mass SMBHs ($\lesssim M_\bullet \sim 10^8 \, \mathrm{M}_\odot$) in relatively low mass galaxies at $z > 4$ has been recently revealed by JWST \citep{2023A&A...677A.145U,2023ApJ...954L...4K,2023arXiv230311946H, matthee2023little}. This population of over-massive MBHs, which is also detected at intermediate red-shifts \citep{overmassiveobs}, significantly exceeds the expectation from the $M_\bullet-\sigma_{*}$ relation \citep[e.g.][]{2023arXiv230812331P}. These new observations challenge our understanding of possible formation and growth mechanisms that have been discussed over the past decades (e.g. \citealp{1984ARA&A..22..471R, 2010A&ARv..18..279V,2023arXiv230812331P}, see \citealp{2023MNRAS.521..241V} for possible observational biases).

The origin of low-mass MBHs, which are often seen as the seeds of SMBHs, is unclear and several, very different, formation mechanisms have been discussed in the literature. One potential origin for MBHs at high-redshift is the remnants of the first stars (PopIII) that are expected to form in $\sim10^{5} - 10^{6} \, \mathrm{M}_\odot$ dark matter halos starting at $z\sim 40$ \citep{2001ApJ...550..372F, Madau_2001,2002ApJ...571...30S,Hirano_2014,2019MNRAS.483.3592B}. Since these seeds have only a small mass $10 \lesssim M_\bullet / \mathrm{M}_\odot \lesssim 1000 $, efficient growth through BH mergers or accretion at super-Eddington rates would be necessary to explain the observed high-redshift quasars in this scenario \citep[e.g.][]{Haiman_2004}. Mechanisms such as the direct collapse of gas in the high-redshift Universe require special environments but can produce MBH seeds with masses of up to $10^5-10^6 \, \mathrm{M}_\odot$ \citep[e.g.][]{2006MNRAS.370..289B,2008ApJ...686..801O, 2014ApJ...795..137R, 2017NatAs...1E..75R, bogdan2023evidence, natarajan2023detection} and hence circumvent the growth timescale problem. Some studies even predict direct collapse BHs with masses of up to $10^8 \, \mathrm{M}_\odot$ \citep{mayer2023direct}. Alternatively, the runaway stellar growth in dense star clusters might produce MBHs with masses of  $10^3-10^4 \, \mathrm{M}_\odot$ \citep[e.g.][]{2002ApJ...576..899P,2009ApJ...694..302D, 2015MNRAS.451.2352K,2011ApJ...740L..42D, 2014MNRAS.442.3616L,Rizzuto_2023}. Another potential origin are primordial black holes which might have formed in the primordial Universe, possibly with a wide range of masses up to the supermassive regime \citep{Carr_2022}. We refer to \citet{Inayoshi_2020} for a detailed review of possible astrophysical formation scenarios. 

These seeding mechanisms make different predictions for the MBH seed mass functions and occupation fractions \citep[e.g.][]{2008MNRAS.383.1079V,2020ARA&A..58..257G}. However, in the hierarchical picture of the growth of cosmological structures, all mechanisms inevitably predict that multiple of these BH seeds can be transported into the same dark matter halos during mergers of halos and galaxies. Furthermore, MBHs (or SMBH "seeds") do not necessarily have to form in the most massive halos, where they are observed today, but have to sink to the halo centres where they can efficiently grow through gas accretion or mergers with other MBHs. 

In a cosmological context, it might be difficult to transport ("seed") MBHs to the centres of dark matter halos and their galaxies. Furthermore, the expected multiple dynamical interactions make it difficult for MBHs to merge and/or result in dynamical ejections from the center. In particular low mass seed MBHs might not easily sink to the galactic centers and continue orbiting in galactic halos \citep[e.g.][]{2003MNRAS.340..647I}. Even if the MBHs sink and merge, gravitational recoil kicks might eject the MBHs. Considering these processes, a population  'wandering' MBHs has been predicted from semi-analytical models \citep{2003ApJ...593..661V,Volonteri_2005}. Studies with cosmological zoom-in simulations by \citet{Ma_2021} find that MBH seeds  with masses $< 10^8 \, \mathrm{M}_\odot$ cannot efficiently migrate to the galaxy centres at high-redshift. Similar results are reported from other zoom-in simulations \citep[e.g.][]{Pfister_2019, 2021MNRAS.505.5129B} as well as larger scale cosmological simulations, e.g. NewHorizon \citep{horizonagn} and Romulus \citep{Tremmel_2017}. These simulations found part of the the predicted population of "wandering black holes" that are free-floating and cannot sink to the halo/galaxy centres. However, these simulations did not take into account dynamical ejection process in the center which, according to \citet{2003ApJ...593..661V,Volonteri_2005}, are very important. A large fraction of the population of these wandering BHs is expected to be hidden due to their low accretion luminosities \citep{2002ApJ...571...30S,2003MNRAS.340..647I,2022ApJ...936...82S}. So far, only a few observations report the detection of off-center BHs \citep[e.g.][]{Mezcua_2020} in low mass halos as well as massive galaxies \citep[e.g.][]{meyer2023alma}. 

It is important to know how and whether MBHs at the different masses predicted by the various proposed formation scenarios can sink to the centres of low-mass halos and galaxies, stay there and merge on short timescales. This determines the MBHs' ability to rapidly grow already at high-redshift into the traditional SMBH regime, as seen by the observations discussed above. The favourite process of almost all successful cosmological models for this growth phase is gas accretion in the centres of galaxies \citep[see e.g.][for reviews]{2015ARA&A..53...51S,2017ARA&A..55...59N}.

However, simulating the sinking and interaction of MBHs in galaxy simulations is a complicated dynamical problem that usually requires simplifications, in particular for the above mentioned cosmological simulations. To resolve dynamical friction, which is responsible for the BH sinking, a mass resolution of the background stellar and dark matter particles has to be significantly higher (higher than a factor of $10^2$, e.g. \citealp{Rantala_2017}) than the MBH seed mass. Furthermore, the force softening between the background particles and the BH must be sufficiently small to account for the point-like nature of the BH \citep[see e.g.][]{Pfister_2019}. Because this is not achievable in typical cosmological simulations, sub-grid prescriptions capturing the effect of dynamical friction on unresolved scales like Chandrasekhar's equation \citep{1943ApJ....97..255C} or more modern formulations \citep{Tremmel_2015, Ma_2023} are employed. These approximate methods are tuned to reproduce the expected sinking timescales, but can not fully capture the dynamical back-reaction of MBHs on their environment. Additionally, the dynamical interaction between MBHs cannot be resolved on small scales such that BHs are often artificially merged on kiloparsec scales. This neglects the important phase of MBH binary or triple evolution during which MBHs dynamically change their environment, can be dynamically ejected, or merge with a subsequent recoil kick. Therefore, most current cosmological simulations cannot make accurate predictions on the sinking, interaction and merging of MBHs \citep[see][for studies to accurately capture MBH interactions in cosmological simulations]{2021ApJ...912L..20M,2022ApJ...929..167M}.  

Many previous studies have demonstrated that coalescing MBHs in the centres of merging galaxies result in the formation of `cores' in the central stellar density profiles. The core formation is mainly caused by the transfer of energy from the MBHs to stars by dynamical friction until the MBHs form a binary. Thereafter slingshot ejections of stars harden the MBHs and further reduce the central density \citep[e.g.][]{2001ApJ...563...34M, 2003ApJ...593..661V, 2005LRR.....8....8M,Rantala_2017,2018ApJ...864..113R,2021MNRAS.508.4610F, 2021MNRAS.502.4794N}. Throughout the paper, we refer to the combined effect of these two processes (dynamical friction heating and slingshot ejections) as black hole "scouring". As shown in \citet{2006ApJ...648..976M}, repeated sinking, binary formation and merger events can enhance this effect, leading to cores in the density distribution and mass deficits in the galactic centres that scale with the MBH mass and the number of MBH sinking events. It has been conjectured that the signature of MBH dynamics could also be imprinted in the dark matter profile \citet{Milosavljevic_2002}. Although not considered in \citet{2006ApJ...648..976M}, the BH merger recoil kick can also have an important effect on the evolution of the host galaxy. In the last phase of the MBH merger, the gravitational wave emission becomes highly anisotropic, resulting in a strong spin and mass ratio dependent velocity kick up to $v_{\rm kick}\sim 5000\, \rm km/s$ \citep{2015PhRvD..92b4022Z, Campanelli_2007}. As shown in \citet{2021MNRAS.502.4794N}, the rapid ejection of the remnant from the galactic centre can unbind the core, adding to the effect of MBH sinking and slingshots. This effect can remove up to $\sim 5 \, \rm {M_\bullet}$ from the central core region of a galaxy \citep{2004ApJ...613L..37B, 2008ApJ...678..780G}. The recoil of merged BHs also has implication for the ability of seeds to grow through mergers, since merger remnants might exceed the escape velocity of their host halos \citep{Haiman_2004, 2006ApJ...650..669V,Volonteri_2005}. Some observational candidates for recoiling BHs are presented in \citet{Caldwell_2014, Chiaberge_2018, offcenterbhs}.

\begin{figure*}
	\includegraphics[width=2\columnwidth]{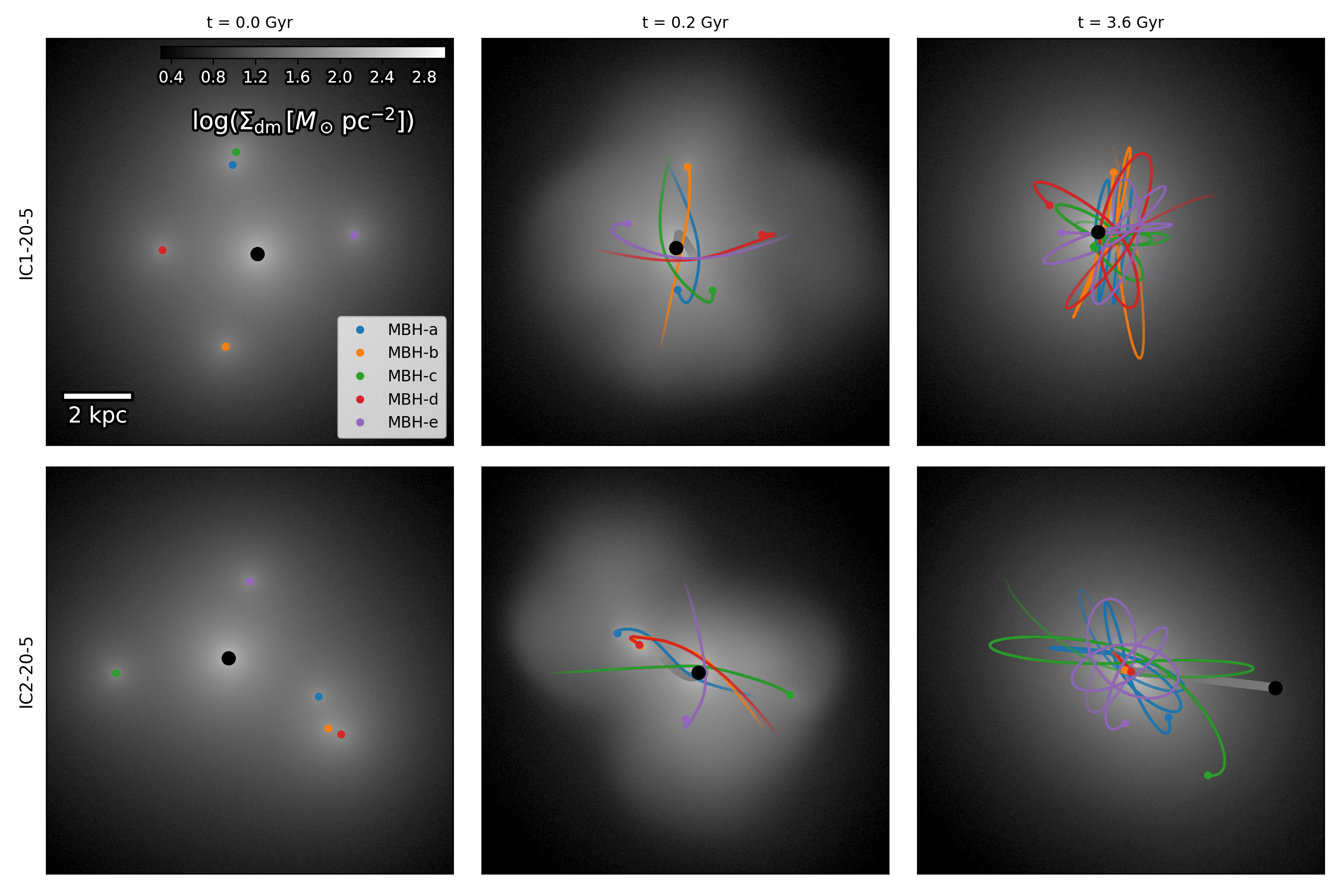}

    \caption{Dark matter surface densities (greyscale) and massive black hole (MBH) orbits (color coded) for the highest resolution ($20 \, \mathrm{M}_\odot$) simulations with orbital configuration {\tt IC1-20-5} (top row) and {\tt IC2-20-5} (bottom row) at initial time (left panels) , after $0.2 \, \rm Gyr$ (middle panel) and after $3.6 \, \rm Gyr$ (right panel). The central halo hosts a MBH with $M_{\bullet, \rm c}=10^5 \, \mathrm{M}_\odot$ while the BHs of the five in-falling satellites only have 20 \% of the central MBH mass each. Interactions of the MBHs with each other and with nearby dark matter particles are calculated accurately with a regularisation method. Some MBHs orbit in the halo at large radii (kpc scales) while MBHs that sink to the centre form dynamically complex subsystems. In the bottom right panel, the central MBH (black) is ejected from the host halo after a merger.}
    \label{fig:initial_conditions}
\end{figure*}

\begin{table*}
    \centering
    \begin{tabular}{l||l|l|l|l|l|l|l|l|l}
       Simulation  & IC & $M_{\rm halo, c}$ [$\mathrm{M}_\odot$] & $M_{\rm *, c}$ [$\mathrm{M}_\odot$] & $\mathrm{log}(M_{\bullet, \rm c}/\mathrm{M}_\odot)$ & $\epsilon_{\rm BH, dm}$ [pc]  & $r_{\rm Ketju}$ [pc] & $m_{\rm res}$ [$\mathrm{M}_\odot$] &  $N_{\rm p}$ \\
 \hline
       
       {\tt IC1-20}           & {\tt 1} & $2 \times 10^9$ & - & ${\tt 3, 4, 5, 6, 7}$ & 0.0    & 12.0 & 20 & $2\times10^8$\\    
       {\tt IC2-20 }          & {\tt 2} & $2 \times 10^9$ & -& ${\tt 3, 4, 5, 6, 7}$ & 0.0    & 12.0 & 20 & $2\times10^8$\\  
       {\tt IC2-20-soft}      & {\tt 2} & $2 \times 10^9$ & -& ${\tt 3, 4, 5, 6, 7}$ & 4.0    & 12.0 & 20 & $2\times10^8$\\       
       {\tt IC2-100  }        & {\tt 2} & $2 \times 10^9$ & -& ${\tt 4, 5, 6, 7}$ & 0.0    & 21.0 & 100 & $4\times10^7$ \\ 
       {\tt IC2-100-stars}    & {\tt 2} & $2 \times 10^9$ & $2 \times 10^8$ & ${\tt 4, 5, 6, 7}$ & 0.0         & 21.0 & 100 & $4.4\times10^7$ \\ 
       {\tt IC2-100-soft-stars}&{\tt 2} & $2 \times 10^9$ & $2 \times 10^8$& ${\tt 4, 5, 6, 7}$ & 7.0      & 21.0 & 100 & $4.4\times10^7$ \\ 
\hline
       {\tt IC1-1000}         & {\tt 1} & $2 \times 10^9$ & -& ${\tt 5, 6, 7}$ & 0.0   & 21.0 & 1000 & $4\times10^6$ \\ 
       {\tt IC2-1000}         & {\tt 2} & $2 \times 10^9$ & -& ${\tt 5, 6, 7}$ & 0.0   & 21.0 & 1000 & $4\times10^6$ \\ 
       {\tt IC3-1000}         & {\tt 3} & $2 \times 10^9$ & -& ${\tt 5, 6, 7}$ & 0.0   & 21.0 & 1000 & $4\times10^6$ \\ 
       {\tt IC4-1000}         & {\tt 4} & $2 \times 10^9$ & -& ${\tt 5, 6, 7}$ & 0.0   & 21.0 & 1000 & $4\times10^6$ \\ 
       {\tt IC1-1000-soft}         & {\tt 1} & $2 \times 10^9$ & -& ${\tt 5, 6, 7}$ & 7.0   & 21.0 & 1000 & $4\times10^6$ \\ 
       {\tt IC2-1000-soft}         & {\tt 2} & $2 \times 10^9$ & -& ${\tt 5, 6, 7}$ & 7.0   & 21.0 & 1000 & $4\times10^6$ \\ 
       {\tt IC3-1000-soft}         & {\tt 3} & $2 \times 10^9$ & -& ${\tt 5, 6, 7}$ & 7.0   & 21.0 & 1000 & $4\times10^6$ \\ 
       {\tt IC4-1000-soft}         & {\tt 4} & $2 \times 10^9$ & -& ${\tt 5, 6, 7}$ & 7.0   & 21.0 & 1000 & $4\times10^6$ \\ 
       
       {\tt IC1-1000-stars}         & {\tt 1} & $2 \times 10^9$ & $2 \times 10^8$& ${\tt 5, 6, 7}$ & 0.0   & 21.0 & 1000 & $4.4\times10^6$ \\ 
       {\tt IC2-1000-stars}         & {\tt 2} & $2 \times 10^9$ & $2 \times 10^8$& ${\tt 5, 6, 7}$ & 0.0   & 21.0 & 1000 & $4.4\times10^6$ \\ 
       {\tt IC3-1000-stars}         & {\tt 3} & $2 \times 10^9$ & $2 \times 10^8$& ${\tt 5, 6, 7}$ & 0.0   & 21.0 & 1000 & $4.4\times10^6$ \\ 
       {\tt IC4-1000-stars}         & {\tt 4} & $2 \times 10^9$ & $2 \times 10^8$& ${\tt 5, 6, 7}$ & 0.0   & 21.0 & 1000 & $4.4\times10^6$ \\ 
       {\tt IC1-1000-soft-stars}         & {\tt 1} & $2 \times 10^9$ & $2 \times 10^8$& ${\tt 5, 6, 7}$ & 7.0   & 21.0 & 1000 & $4.4\times10^6$ \\ 
       {\tt IC2-1000-soft-stars}         & {\tt 2} & $2 \times 10^9$ & $2 \times 10^8$& ${\tt 5, 6, 7}$ & 7.0   & 21.0 & 1000 & $4.4\times10^6$ \\ 
       {\tt IC3-1000-soft-stars}         & {\tt 3} & $2 \times 10^9$ & $2 \times 10^8$& ${\tt 5, 6, 7}$ & 7.0   & 21.0 & 1000 & $4.4\times10^6$ \\ 
       {\tt IC4-1000-soft-stars}         & {\tt 4} & $2 \times 10^9$ & $2 \times 10^8$& ${\tt 5, 6, 7}$ & 7.0   & 21.0 & 1000 & $4.4\times10^6$ \\ 
\hline
    {\tt IC3-1000-perturbed}  & 3 & $2 \times 10^9$ & - & ${\tt 7}$     & 0.0   & 21.0 & 1000 & $4\times10^6$ \\ 
\hline
      {\tt IC1m-1000} & {\tt 1m} & $2 \times 10^{10}$   & -& ${\tt 5, 6, 7}$           & 0.0   & 21.0 & 1000 & $4 \times 10^7$ & \\ 
      {\tt IC1m-1000-stars} & {\tt 1m} &$2 \times 10^{10}$  & $2 \times 10^8$&  ${\tt 5, 6, 7}$    & 0.0       & 21.0 & 1000 & $4.04 \times 10^7$ &\\ 
      {\tt IC1m-1000-soft-stars} & {\tt 1m} &$2 \times 10^{10}$  & $2 \times 10^8$&  ${\tt 5, 6, 7}$     & 7.0       & 21.0 & 1000 & $4.04 \times 10^7$ &\\ 
      {\tt IC2m-1000} & {\tt 2m} &$2 \times 10^{10}$  & - &  ${\tt 5, 6, 7}$           & 0.0   & 21.0 & 1000 & $4 \times 10^7$ & \\ 
      {\tt IC2m-1000-stars} & {\tt 2m} & $2 \times 10^{10}$  & $2 \times 10^8$ &${\tt 5, 6, 7}$     & 0.0       & 21.0 & 1000 & $4.04 \times 10^7$ &\\ 
\hline       
       
    \end{tabular}
    \caption{Summary of the simulation presented in this work. 
    Every simulation consists of a main central halo with a BH and five infalling sattelite halos (each one carries an additional BH with $20\%$ of the mass of the BH in the central halo). In the fiducial case, the central mass is $M_{\rm halo, c} = 2 \times 10^9 \, \rm M_\odot$ and satellites have $M_{\rm halo, s} = 4 \times 10^8 \, \rm M_\odot$. The name of the simulation represents the orbital configuration of the merger ({\tt IC}), mass resolution $m_{\rm res}$ and logarithm of the mass of the central MBH $M_{\bullet, \rm c} = 0.2 M_{\bullet, \rm s}$. We vary the MBH mass is steps of 1 dex. If the force between BH and dark matter is softened (with softening $\epsilon_\mathrm{BH,dm}$) or a stellar component (of mass $M_{*, \rm c}$) is used, we add the keyword "{\tt soft}" or "{\tt stars}". We also list the mass of the central halo $M_{\rm halo, c}$ and add "{\rm m}" for the more massive halos ($M_{\rm halo, c} = 2 \times 10^10 \, \rm M_\odot$) to distinguish them from the fiducial simulations. The size of the regularized region is given by $r_{\rm Ketju}$. The softening among dark matter and star particles ($\epsilon$ = $\epsilon_\mathrm{*,dm}$ = $\epsilon_\mathrm{dm,dm} = \epsilon_\mathrm{*,*}$) is always set to $\epsilon = r_{\rm Ketju}/3$. The total number of particles $N_{\rm p}$ in each simulation is shown in the last column. Simulations in the first section of the table are designed to maximize resolution while the second section aims at improving the statistics with multiple lower-resolution initial conditions. With {\tt IC3-1000-perturbed}, we test the impact of tiny perturbations on the system. The last section lists simulations at a higher galaxy mass, at the expense of resolution. }
    \label{tab:simulations}
\end{table*}

In this paper we aim to overcome technical limitations of cosmological simulations and study the sinking, interaction and merging of MBHs in idealized multiple galaxy mergers. Our high-resolution simulations use accurate unsoftened gravitational forces for interactions of MBHs with each other, with stars, and with dark matter particles. They are designed to accurately capture the effect of dynamical friction as well as the dynamical interaction of MBHs with their dark matter and stellar environment using a regularization technique, post-Newtonian corrections and recoil kicks \citep{Rantala_2017,2018ApJ...864..113R, 2021ApJ...912L..20M, Mannerkoski_2023}. We focus our study on merging low-mass halos and galaxies, resembling the rapid hierarchical growth of structure at high-redshift \citep[see e.g][and references therein]{2022MNRAS.511.4044D}. This allows us to achieve sufficient resolution in the stellar and dark matter components to follow low-mass MBHs (such as predicted by the PopIII seeding scenario). In our idealized merger setup, we can represent dark matter particles and stars with masses as low as $20 \, \mathrm{M}_\odot$ and $100 \, \mathrm{M}_\odot$ respectively, close to the individual star limit. This allows us to follow the dynamics of MBHs as low as $\sim 1000 \, \mathrm{M_\odot}$.

With our simulations, we gain a better understanding of whether and how MBHs at different masses can sink to the galaxy/halos centres which is a long standing question in particular for low mass seed MBHs \citep[e.g.][]{2002ApJ...571...30S}. We also investigate whether the forming binary or multiple MBH systems result in mergers. As these processes are expected to be MBH mass dependent, we cover a wide range of masses from the PopIII seed to the direct collapse seed regime. For low-mass halos, the escape velocities are low, such that dynamical ejections of MBHs as well as recoil ejections of MBHs merger remnants are possible. With our simulations, we determine typical scenarios with implications for the growth of MBHs through mergers. In addition, we study how the presence of sinking and interacting MBHs impacts the stellar and dark matter density distribution.

The paper is structured as follows: We introduce our simulation framework {\sc Ketju} and the initial conditions in section \ref{sec:ketju_method}. In section \ref{sec:dm}, we discuss simulations with dark matter halos with a focus on sinking (\ref{sec:dm_sink}), binary and triple formation (\ref{sec:dm_semi}) and impact on density distributions (\ref{sec:dm_dens}). We repeat the discussions for simulations with a galaxy (i.e. including a stellar component) in the dark matter halos in section \ref{sec:dmstar}. After a discussion of our results in section \ref{sec:discussion}, we conclude in \ref{sec:conlusion}.

\section{Simulations with Ketju}
\label{sec:ketju_method}
To accurately follow the dynamics of BHs and the interaction with their stellar and dark matter environment, we perform our numerical experiments with the simulation code {\sc Ketju}. The {\sc Ketju} code project is an extension of the tree gravity solver {\sc Gadget-3} and was introduced in \citet{Rantala_2017,2018ApJ...864..113R, 2021ApJ...912L..20M}. Although not used here yet, an updated version using {\sc Gadget-4} was recently presented in \citet{Mannerkoski_2023}. We give a basic description of the methods in the following section.

\subsection{Ketju}

In regions close to BHs (the "{\sc Ketju} region" with radius $r_{\rm Ketju}$, see Tab. \ref{tab:simulations}), {\sc Ketju} switches from the standard {\sc Gadget-3} leapfrog integrator to the accurate {\sc MSTAR} integrator based on algorithmic regularization \citep{2020MNRAS.492.4131R}. 

The regularized integrator relies on three basic ingredients to achieve high integration accuracy \citep[e.g.][]{1999MNRAS.310..745M,1999AJ....118.2532P,2006MNRAS.372..219M,2008AJ....135.2398M}. First, time transforming the equations of motion together with the use of the common leapfrog integrator circumvents the Newtonian singularity at small particle separations. Second, the use of a minimum spanning tree (MST) inter-particle coordinate system significantly reduces the numerical floating-point round-off error. Finally, the Gragg–Bulirsch–Stoer (GBS) extrapolation method \citep{1965SJNA....2..384G,Bulirsch1966} guarantees the desired integration accuracy for all the dynamical variables of the system, controlled by the relative GBS tolerance parameter $\eta_{\rm GBS}$. 

The regularization scheme allows for solving the orbit of the Newtonian two-body problem at in principle machine precision. MSTAR is seamlessly integrated in the {\sc Gadget-3} time integration and gravity solver. The regularized "Ketju region" that is carried around by each BH can merge with the regularized region of another BH particle if they overlap. To allow for a smooth transition of particles in- and out of the Ketju region, the gravitational softening $\epsilon$ of the simulation particles must be sufficiently (at least by a factor of $2.8$) smaller than the size of the Ketju region $r_{\rm Ketju}$. The perturbations between the regions and the more distant particles in {\sc Gadget-3} are performed using a second-order Hamiltonian splitting technique. Hence, {\sc Ketju} can follow the orbits of simulation particles at close separations around BHs without gravitational softening, without suffering the typical large errors during close encounters between particles in simulations. The regions are integrated in parallel in an efficient manner such that particle numbers of several thousand can be accurately integrated in each Ketju region. 

The interaction between BHs is computed using the post-Newtonian equation up to order 3.5, allowing to track the gravitational wave driven coalescence of black holes. BHs merge if they approach closer than $10$ Schwarzschild radii $R_s$. Based on the mass ratio and spin parameters of the merger progenitors, the merger remnant receives a recoil kick based on the small-scale GR-simulations presented in \citep{2015PhRvD..92b4022Z}.

\subsection{Initial conditions}
\label{sec:ics}
We set-up multiple merger initial conditions for low mass galactic dark matter halos with a central halo and five smaller satellite halos on bound orbits. The energies of the satellite orbits are chosen such that the entire system merges within a few Gyr. Under this condition, we have realised four random initial configurations with different orbits (positions and velocities) for the satellite galaxies ({\tt IC1}, {\tt IC2}, {\tt IC3}, {\tt IC4}). This allows us to investigate idealised multiple dark matter halo, galaxy and BH interactions at reasonable computational cost. Such an idealised multiple merger scenario for low-mass galaxies might be representative of high-redshift environments ($z \gtrsim 2$), where theoretical models predict that low mass halos have undergone several minor merger events within a few Gyr \citep[see e.g][and references therein]{2022MNRAS.511.4044D}. 

Our fiducial simulations have dark matter halos with radial densities following a \citet{1990ApJ...356..359H} profile with $r_{\rm dm,{1/2}} = 7.1 \, \rm kpc$ and masses of $M_{\rm dm} = 2 \times 10^9 \, \mathrm{M}_{\odot}$ (central) and $M_{\rm dm} = 4 \times 10^8 \, \mathrm{M}_{\odot}$ (satellites). In some simulations, the halos host a central galaxy with a stellar mass of $0.1 \times M_{\rm dm} $ and a half-mass radius $r_{\rm *, {1/2}} = 1 \, \rm kpc$ (labeled "{\tt -star}" in Tab. \ref{tab:simulations}), also following a Hernquist profile. This baryonic mass fraction is higher than predictions from e.g. abundance matching and empirical galaxy formation models \citep{2013MNRAS.428.3121M,2018MNRAS.477.1822M,2019MNRAS.488.3143B,2023MNRAS.520..897O}. Therefore, the simulations place an upper limit on the supporting effect of a stellar component on the dynamical friction to speed up the galaxy and BH merger process, even though the baryon fractions of dark matter halos at high-redshift are still highly uncertain. To test the effect of a higher halo mass, we also perform simulations at ten times higher mass, i.e. a central halo mass of $M_{\rm dm} = 2 \times 10^{10} \, \mathrm{M}_{\odot}$ ($r_{\rm dm,{1/2}} = 16.9 \, \rm kpc$) and satellite halo mass of $M_{\rm dm} = 4 \times 10^9 \, \mathrm{M}_{\odot}$ ($r_{\rm dm,{1/2}} = 7.1 \, \rm kpc$). The mass of the galaxy - if present - has the same mass and size as in our fiducial simulations. These simulations are labeled {\tt IC1m} and {\tt IC2m} in Tab. \ref{tab:simulations} and have orbits similar to {\tt IC1} and {\tt IC2} respectively, although scaled to higher masses.

All halos are initialized in equilibrium with their BHs with the technique presented in \citet[][see also \citealp{2012MNRAS.425.3119H}]{Rantala_2017}. We study the evolution of massive BHs (MBHs) in a mass regime above stellar BHs from 200 to $10^7 \, \mathrm{M}_\odot$. Traditionally, these objects are termed intermediate mass BHs (IMBHs with $200 \, \mathrm{M}_\odot \lesssim M_{\bullet} \lesssim 10^5 \, \mathrm{M}_\odot$) and supermassive BHs (SMBHs with $M_\bullet \gtrsim 10^6 \, \mathrm{M}_\odot$). The satellite galaxies also host central MBHs with masses of 20\% of the MBH in the main galaxy ($M_{\bullet, \rm s} = 0.2 \times M_{\bullet, \rm c}$). With recent observations indicating the existence of over-massive BHs in low mass galaxies both at lower redshifts \citep{overmassiveobs} and also at early cosmic times \citep{2023A&A...677A.145U,2023arXiv230311946H} as well as the uncertainties of the seeding mechanisms \citep[see e.g.][]{2020ARA&A..58..257G}, it is crucial to consider a wide range of BH masses beyond the traditional stellar mass - BH mass relation in the local Universe \citep{2013ARA&A..51..511K}. All MBHs are assumed to have zero spin initially.

We run simulations with three different mass resolutions for the dark matter particles ($20, 100$, and $1000 \, \mathrm{M}_\odot$ with labels "{\tt 20}", "{\tt 100}" and "{\tt 1000}", respectively) resulting in particle numbers of $\sim10^6 - 10^8$. The gravitational interactions of the dark matter particles with each other are softened on a scale of $\epsilon =4 \, \rm pc$ for the highest resolution runs ($m_\mathrm{dm}=20 \, \mathrm{M}_\odot$) and $\epsilon =7 \, \rm pc$ for the lower resolution simulations ($m_\mathrm{dm}=100 \, \mathrm{M}_\odot$ and $m_\mathrm{dm}=1000 \, \mathrm{M}_\odot$). The fiducial runs allow for unsoftened (collisional) interactions of BHs with the dark matter particles inside the Ketju region of $r_{\rm Ketju} = 3 \, \epsilon$, i.e. 12 pc at the highest resolution (e.g. simulation {\tt IC1-20}) and 21 pc at lower resolution (e.g. {\tt IC1-100} and  {\tt IC1-1000})\footnote{To avoid the stalling of the integration in three simulations  with a extremely small binary separation and large particle numbers is the Ketju region, we decreased the size of the Ketju region to $r_{\rm Ketju} = 3\, \rm pc$ ({\tt IC1-20-4}, at $2.2 \rm \, Gyr$) and $r_{\rm Ketju} = 12 \, \rm pc$ ({\tt IC2-100-star-4}, at 6 Gyr). This also decreases the global softening to $\epsilon_{\rm soft} = r_{\rm Ketju}/3$.}.
The unsoftened dark matter BH interactions allow for the hardening and merging of bound BH subsystems due the interactions with dark matter particles. We also perform comparison runs with softened interactions of the dark matter with the back holes (i.e. simulation name {\tt IC1-20-soft} for the highest resolution simulation). The central stellar component, if present, has the same masses and softening lengths as the dark matter to reduce additional relaxation effects and the interactions of stars with BHs are always unsoftened. We have run simulations including a central stellar component at $100 \, \mathrm{M}_\odot$ and 1000 $\mathrm{M}_\odot$ resolution (i.e. simulation name {\tt IC2-100-stars} and {\tt IC2-1000-stars}, respectively). For comparison, all simulations are also performed without MBHs. The suite of over 100 simulations is summarized in Table \ref{tab:simulations}. While the simulations in the first section of the table aim at the highest possible mass resolution, the second section is designed to test multiple realizations. With initial conditions {\tt IC3-1000-perturbed}, we test the impact of small perturbations on the dynamics of the system. The last section of the table represents the simulations with higher halo mass. Throughout the paper, we will refer to simulations in this table using the label given in the first column together with the extension "{\tt 3, 4, 5, 6, 7}" indicating the logarithm of the central MBH mass $M_{\bullet, \rm c}$ in the particular simulation.

In Fig. \ref{fig:initial_conditions}, we show an example of the highest resolution initial configuration of {\tt IC1-20-5} (top row, left) and {\tt IC2-20-5} (bottom row, left) with the orbits of the MBHs color coded after 0.2 Gyr (middle panels) and after 3.6 Gyr when the merging of the dark matter holes is complete and the system has relaxed (right panels). As indicated by the extension "{\tt -5}" in the label, the presented simulations have a central MBH mass of $10^5 \, \mathrm{M}_\odot$.

\begin{figure*}
	\includegraphics[width=2\columnwidth]{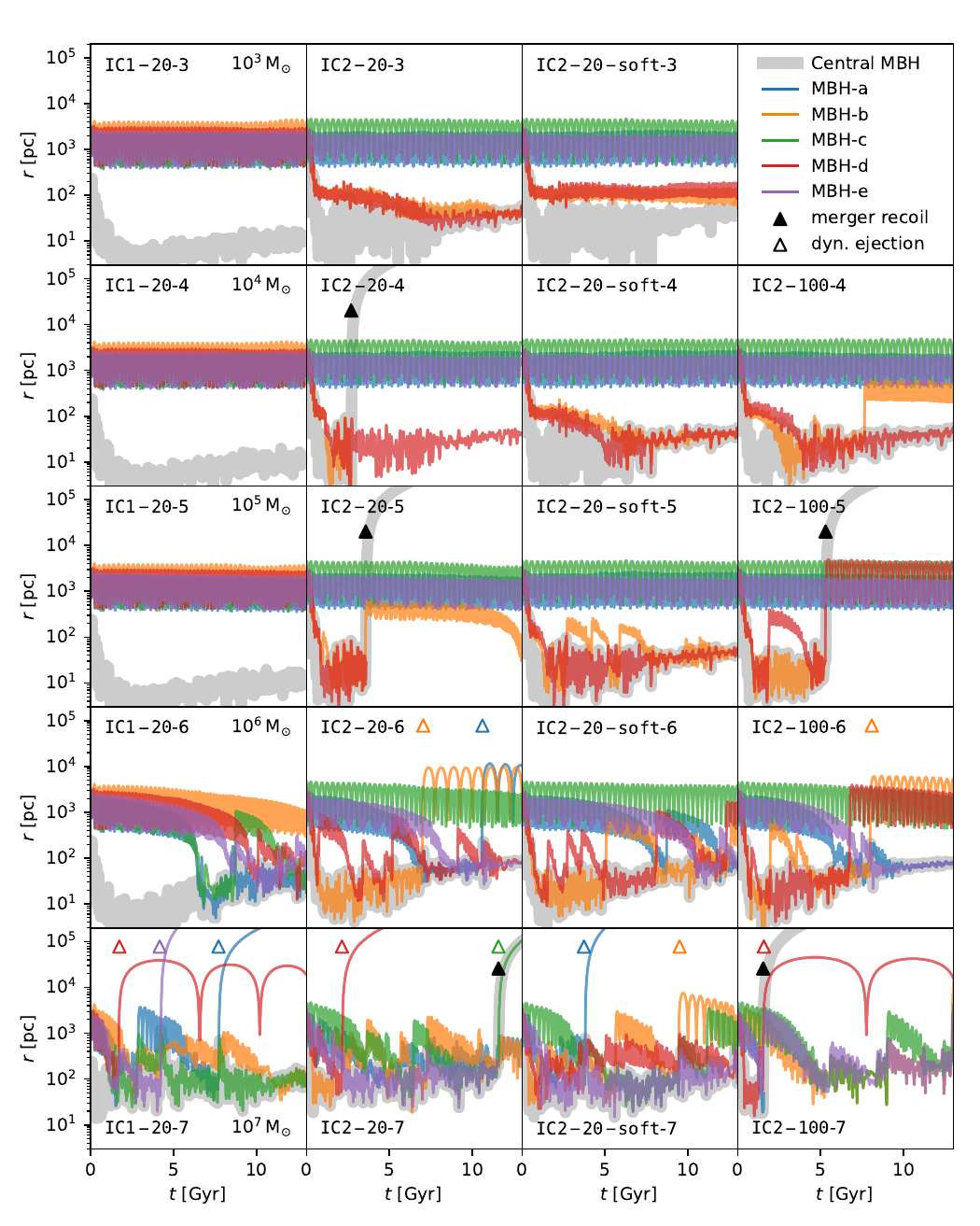}
    \caption{Radial distances of all MBHs (colored lines) from the dark matter density centre of the systems as a function of time. From top to bottom, the central MBH mass increases by factors of 10 from $M_{\bullet, \rm c}=10^3 \, \mathrm{M}_\odot$ to $10^7 \, \mathrm{M}_\odot$ (indicated by the grey band). The two left columns show the highest resolution runs {\tt IC1-20} and {\tt IC2-20}. In the third column, we show the effect of softened dark matter - MBH interactions ({\tt IC2-20-soft}). The right column shows the lower-resolution simulation {\tt IC2-100}. If satellite MBHs sink to the centre, they can form bound systems with the central MBH. They can merge followed by recoil kick ejection (i.e. {\tt IC-20-5}, filled triangle), be dynamically kicked to wide halo orbits without subsequent sinking (i.e. MBH-b, orange, and MBH-a, blue, in {\tt IC2-20-6}) and with subsequent sinking (i.e. MBH-d, red, in {\tt IC2-20-6}), or be dynamically ejected from the system (i.e. MBH-e, purple,and MBH-a, blue, in {\tt IC1-20-7}, open triangles). Low-mass satellite MBHs typically cannot sink to the halo centre (e.g. {\tt IC1-20-3}, {\tt IC1-20-4}, {\tt IC1-20-5}).}
    \label{fig:sinking_dm}  
\end{figure*}

\begin{figure*}
	\includegraphics[width=2\columnwidth]{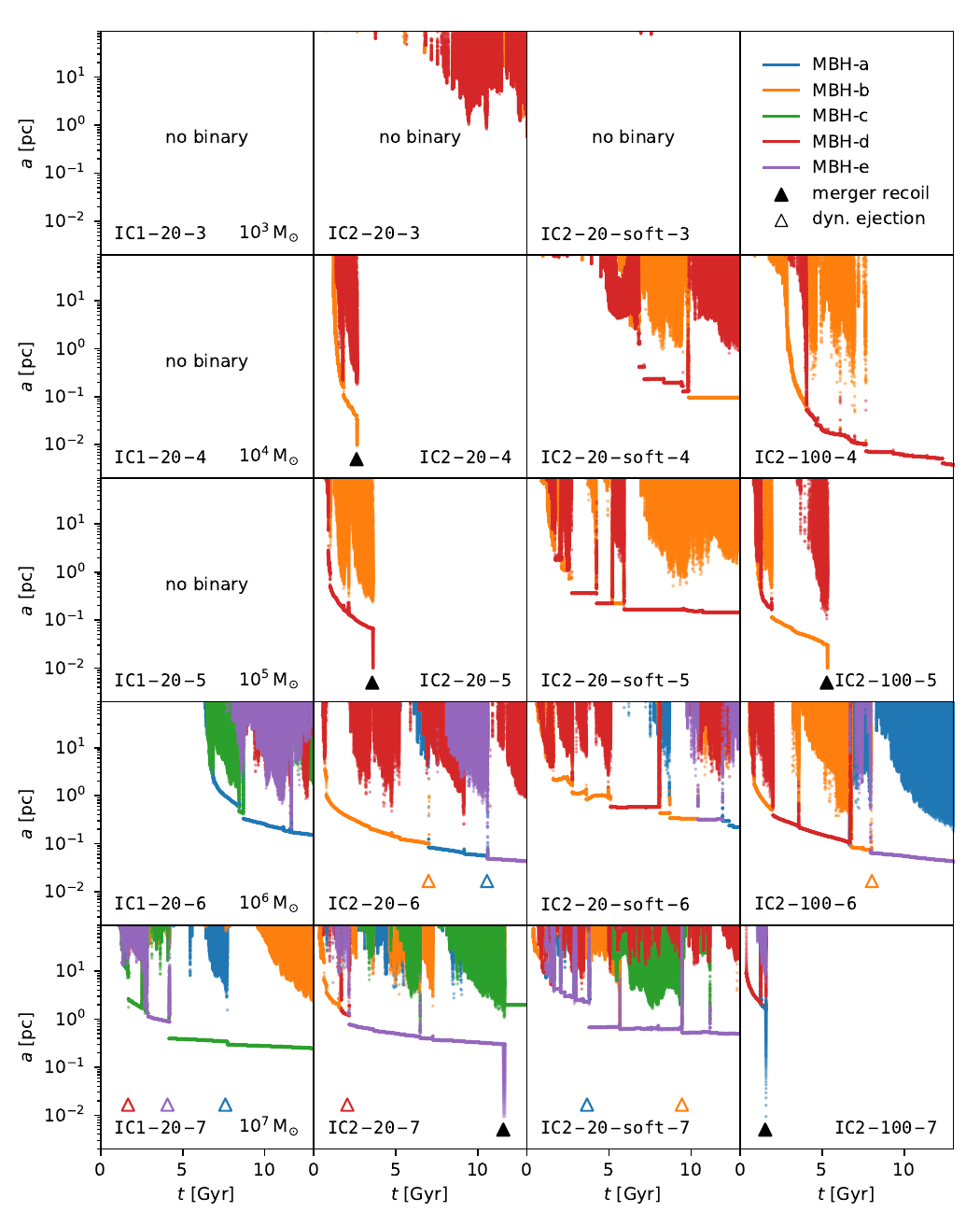}
    \caption{Semi-major axes, $a$, of the satellite MBHs and the more massive host MBHs as a function of time. Panels are arranged as in Fig. \ref{fig:sinking_dm}. Empty panels indicate no MBH binary formation. In gravitational wave emission driven mergers, the semi-major axes rapidly drop to zero (i.e. {\tt IC-20-6}, filled triangles). For collisional (unsoftened) MBH - dark matter interactions, the binaries harden by interactions with the dark matter particles and regularly exchange and kick satellite MBHs in three-body interactions. For example in {\tt IC-20-6}, MBH-b (orange) is replaced by MBH-a (blue), which is replaced by MBH-e (purple). All replaced MBHs are ejected. With force softened MBH - dark matter interactions the binary MBHs do not harden and $a$ only shrinks through BH-BH encounters (i.e. {\tt IC2-20-soft-5}). Dynamical ejections (see e.g. {\tt IC2-20-6}) typically happen at $0.05 - 1 \, \rm pc$ semi-major axes (depending on the MBH mass scale) and are indicated by open triangle (as in Fig.\ref{fig:sinking_dm}).}
    \label{fig:semimajor_axis}
\end{figure*}

\begin{figure}
	\includegraphics[width=1\columnwidth]{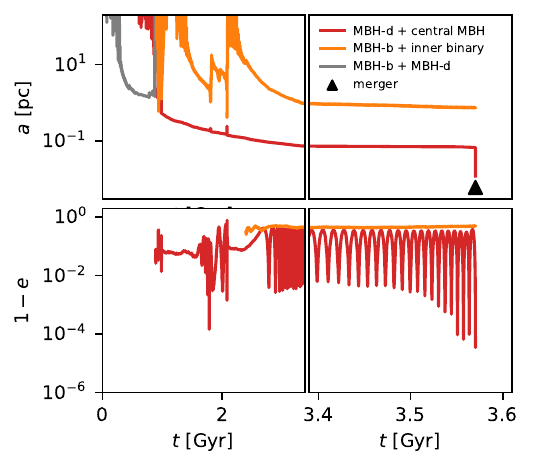}
	\includegraphics[width=1\columnwidth]{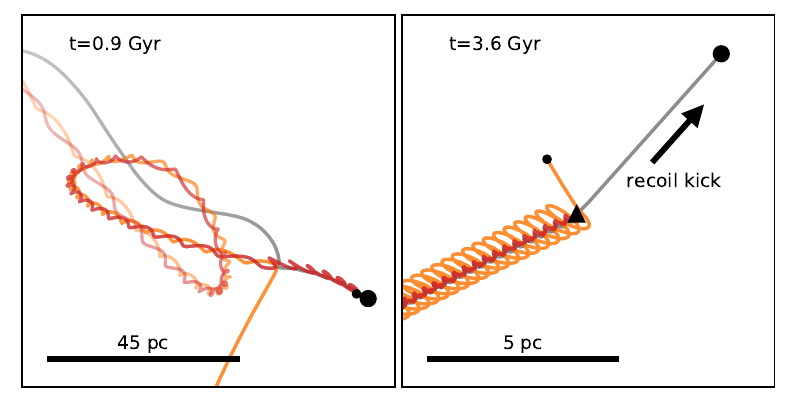}
    \caption{The time evolution of semi-major axes (top panel),  eccentricities (middle panel), and orbits (bottom panel) at the time of triple formation (left) and the gravitational wave driven merger (right, at higher time resolution) for {\tt IC2-20-5}. The central MBH (grey) and two satellite MBHs (MBH-d, red; MBH-b, orange) form a hierarchical triple with von Zeipel-Lidov-Kozai oscillations in the eccentricity evolution (MBH-d, red) starting at $\sim$ 2.5 Gyr. The central binary eccentricity increases until the two MBHs merge at $\sim 3.57$ Gyr and the remnant is ejected (grey line, bottom right panel). The third, originally least bound, MBH (MBH-b, orange) becomes unbound and leaves the centre into a wide halo orbit (see {\tt IC2-20-5} in Fig. \ref{fig:sinking_dm}). The circularisation of the merging MBH binary orbit by gravitational wave emission is captured by the PN corrections in the code but the circularisation timescale is too short to be visualised here (see Fig. \ref{fig:ZLK_stars} for an example).}
    \label{fig:merger_ejection}
\end{figure}
\begin{figure*}
	\includegraphics[width=2\columnwidth]{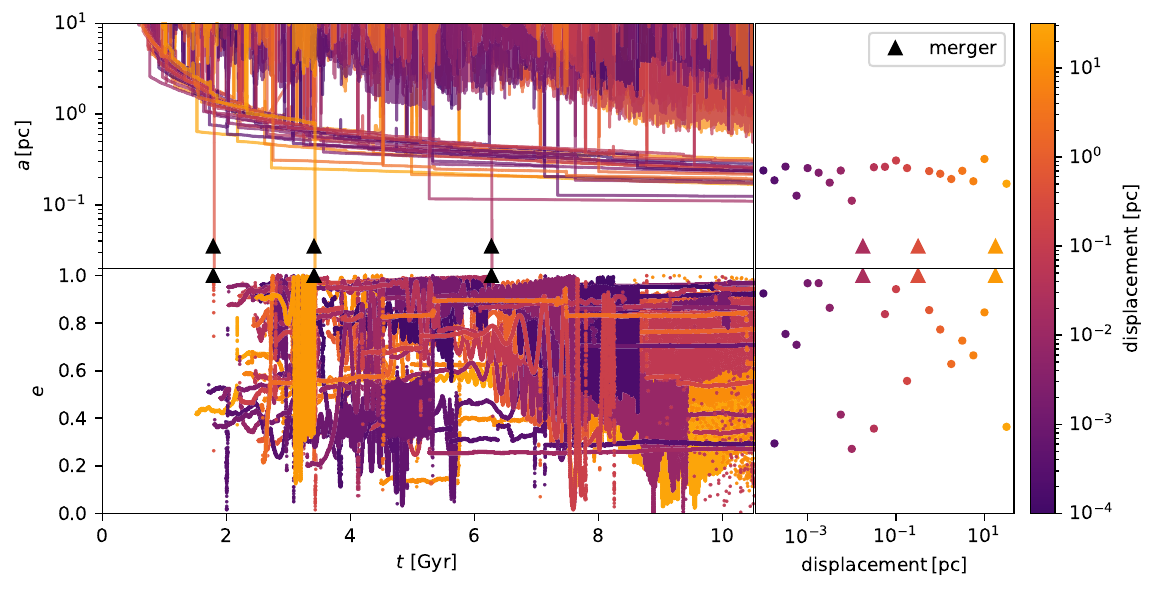}
    \caption{The evolution of the semi-major axes (top left) and the eccentricities (bottom left) of the forming central MBH binaries for 23 different realisations of {\tt IC3-1000-perturbed} (different colours). Initially, the central MBHs were displaced in x-direction by dx $ = 10^{-4} - 30$ pc. All simulations show a similar hardening of the binary with similar final semi-major axes, independent of the initial displacement (bottom left and right). The eccentricities show a stochastic behaviour which, for three simulations with the highest eccentricities, result in MBH mergers (semi-major axes dropping to zero in the top left panel, filled triangles in the right panels). There is no correlation of initial displacement with final binary eccentricity (bottom right panel) and most realisations do not result in a MBH merger within a Hubble time. The particular satellite MBH, which forms a binary with the central MBH, also changes with each realisation.}
    \label{fig:butterfly_a}
\end{figure*}

\section{Massive black holes in merging dark matter halos}
\label{sec:dm}

In this section, we are presenting the mergers of dark matter halos without a stellar component and analyse MBH sinking, binary formation, mergers and the impact on the density structure of the halo.

\subsection{Massive black hole sinking}
\label{sec:dm_sink}
In Fig. \ref{fig:sinking_dm}, we show the distance of the satellite MBHs and the five times more massive host MBH to the region with the highest dark matter density as determined with a shrinking spheres algorithm \citep{power_2003} for 13 Gyr of evolution. The columns represent different high-resolution initial conditions as listed in Tab. \ref{tab:simulations}. Here, we consider two different merger configurations ({\tt IC1} and {\tt IC2}) at two different resolutions ($20\, \mathrm{M}_\odot$ and $100 \, \mathrm{M}_\odot$) and highlight the impact of softened dark matter-MBH forces ({\tt IC2-20-soft}) compared to the fiducial unsoftened dark matter-MBH forces ({\tt IC2-20}). The mass of the satellite MBHs (solid colored lines) increases from $M_{\bullet, \rm s} = 200 \, \mathrm{M}_{\odot}$ in the top row to $M_{\bullet, \rm s} = 2 \times 10^6 \, \mathrm{M}_{\odot}$ in the bottom row.\footnote{For visualization purposes, we have convolved the lines in Fig. \ref{fig:sinking_dm} and \ref{fig:sinking_stars} with a window function to suppress oscillations with a period below $\sim 50 \, \rm Myr$.}

As the first column shows ({\tt IC1-20}), low mass MBHs (first, second and third row) generally do not sink to the halo centre efficiently. Satellite MBHs of mass $M_\mathrm{\bullet,\rm s} \gtrsim 2 \times 10^5 \, \mathrm{M}_{\odot}$ (central MBH mass of $M_\mathrm{\bullet,\rm c} \gtrsim 10^6 \, \mathrm{M}_\odot$, fourth row) start to sink on a timescale shorter than the Hubble time, but not all have reached the center by the end of the simulation. Only in the most massive case, all satellite MBHs ($M_{\bullet, \rm s} \gtrsim 2\times 10^6 \, \mathrm{M}_\odot)$ sink within $\sim2 \rm \, Gyrs$.

The initial conditions {\tt IC2} (column two, three and four) favour the sinking of at least two MBHs as two halos (and their MBHs) are on a very radial initial orbit and rapidly form a bound subsystem (orange and green trajectories in Fig. \ref{fig:initial_conditions}). As a result of this orbital configuration, the satellite MBHs "b" and "d" migrate to the central region, irrespective of their mass. In simulations with a central MBH mass of $10^3-10^5 \, \mathrm{M}_\odot$, only these two satellite MBHs reach the centre of the halo. For central MBH masses $\ge 10^4 \, \mathrm{M}_\odot$, these three MBHs form triple black holes systems that can result in the merger of a satellite MBH (i.e. MBH-b, orange, {\tt in IC2-20-4} and MBH-d, red, in {\tt IC2-20-5}) and the grey host MBH (black triangle). As indicated by the rapidly increasing radial distance of the merger remnant (thick grey line), the MBH merger recoil kick velocity of $v_{\rm kick} \sim 134 \, \mathrm{km/s}$ exceeds the escape velocity of the host halo, unbinding the merger remnant from the host halo (i.e. at about $\sim2.5$ Gyr and $\sim4$ Gyr in {\tt IC2-20-4} and {\tt IC2-20-5}, respectively).

Additional MBHs that are bound to the central MBH binary are often dynamically ejected from the centre at the time of the merger (i.e. MBH-b, orange, in {\tt IC2-20-5}). As an example, the details of the merger process in simulation {\tt IC2-20-5} are shown in Fig. \ref{fig:merger_ejection} and will be discussed in more detail in the next section. A MBH merger and ejection happen at both tested resolutions for a central MBH mass of $10^5 \, \mathrm{M}_\odot$ ({\tt IC2-20-5} and {\tt IC2-100-5}). In the lower resolution case, however, MBH-b and MBH-d are interchanged such that the central MBH merges with MBH-b (orange) and MBH-d (red) is ejected. For the smaller MBH mass, we only find a merger at the higher resolution {\tt IC2-100-4}, although the sinking in the first Gyr of both simulations is very similar. 

For satellite MBHs of mass $2 \times 10^5 \, \mathrm{M}_\odot$ (models '{\tt -6}', fourth row in Fig. \ref{fig:sinking_dm}) more MBHs sink to the centre resulting in complex central MBH subsystems but no MBH mergers. More massive satellite MBHs of $2 \times 10^6 \, \mathrm{M}_\odot$ (models '{\tt -7}', fifth row in Fig. \ref{fig:sinking_dm}) interact so strongly that up to three satellite MBHs can be ejected through tree body encounters (e.g. {\tt IC1-20-7}, open triangles). If the dynamical kicks are weak, the kicked MBHs rapidly sink to the centre again (e.g. {\tt IC2-20-6}, MBH-d, red). Stronger interactions kick MBHs to wide orbits, still bound to the system (e.g. {\tt IC2-20-6}, MBH-b, orange, and MBH-a, blue; {\tt IC1-20-7}, MBH-d, red). The strongest interactions unbind the kicked MBHs from the halo (e.g. {\tt IC1-20-7}, MBH-e, purple, and MBH-a, blue). In the plots throughout the paper, any N-body encounter that kicks satellite MBHs to distances larger than $5 \, \rm kpc$ is labeled "dynamical ejection" and is indicated by an open triangle. These dynamical interactions are often associated with an exchange of the most bound MBH at the centre (see e.g. the orange MBH-b in {\tt IC2-20-6}). This will be discussed in more detail in the context of the semi-major axis evolution. 

In the low resolution case {\tt IC2-100-7}, the central MBH and a satellite MBH merge (black filled triangle) within the first two Gyr and remnant is ejected. In the high resolution version of the same initial conditions ({\tt IC2-20-7}), a merger with a different satellite MBH ejects the host MBH significantly later after $\sim 12 \, \rm Gyr$. Because these simulations are in the resolution regime where dynamical friction and BH binary scouring are sufficiently resolved (mass ratios of $10^5$ and $2 \times 10^4$ respectively), this could be an indication of the stochastic nature of the merger process. Due to the partially chaotic nature of many-body interactions, it is unclear if exact agreement or convergence for runs with different resolutions can in general be achieved \citep[see e.g.][and references therein]{2020MNRAS.497..739N,2023arXiv230708756R}. We discuss some aspects of this behaviour in the next chapter. 

Away from the centre of the halo, where the MBHs have not interacted with the central binary yet, the sinking timescales do not change with resolution indicating that dynamical friction is sufficiently well resolved (see e.g. {\tt IC2-20-6} and {\tt IC2-100-6} for MBH-a (blue) and MBH-b (purple) in Fig. \ref{fig:sinking_dm}). This has also been shown in previous studies including regularised integration around sinking BHs \citep[see e.g.][]{2015MNRAS.452.2337K,Rantala_2017}. The sinking timescales in the model with softened BH-dark matter interactions (e.g. {\tt IC1-20-soft-6}) are only slightly longer than in the unsoftened case because the softening scale of 4 pc is still very small. The small force softening is necessary because dynamical friction cannot be accurately captured anymore for lower resolution and larger force softening length \citep[see e.g.][]{Pfister_2019}.

\subsection{Massive black hole binary formation, triple interactions, and mergers}
\label{sec:dm_semi}
In Fig. \ref{fig:semimajor_axis}, we show the time evolution of the semi-major axes of the host MBH with respect to the satellite MBHs once they have formed a bound system.\footnote{Orbital elements are calculated including third order PN corrections according to \cite{Memmesheimer_2004}. We neglect the potential of the environment and compute the pairwise orbital elements for each combination separately.} The figure is arranged as Fig. \ref{fig:sinking_dm} and empty panels indicate that no central binary has formed. 

Small semi-major axes $a$ indicates that the central MBH is gravitationally bound with the satellite MBH of the respective color, assuming that the potential of stars and dark matter can be neglected on small scales. Solid lines indicate the formation of tight binaries while colored surfaces show the presence of an outer triple companion of the central binary (e.g. the blue surface in {\tt IC2-100-6}). In the latter case, the semi-major axis fluctuates because of the high orbital velocity of the central MBH in the inner binary.\footnote{It is in principle possible to correct for the orbital velocity of the inner binary. In Figure \ref{fig:merger_ejection}, we compute the orbital elements with respect to the centre of mass properties of the inner binary such that the oscillations in the semi-major axis of the outer triple MBH disappear (e.g. the orange surface of {\tt IC2-20-5} in Fig. \ref{fig:semimajor_axis} becomes a solid line in Figure \ref{fig:merger_ejection}).}  Whenever the semi-major axis is rapidly dropping towards zero the host MBH is merging with a satellite MBH (i.e. {\tt IC2-20-5} or {\tt IC-2-100-7}, filled black triangles). 

As already discussed in Fig. \ref{fig:sinking_dm}, for {\tt IC1} (first column), only the most massive MBHs sink fast enough to form a binary within a Hubble time ({\tt IC1-20-6} and {\tt IC1-20-7}). As a result of the initial orbital configuration in {\tt IC2}, two satellite MBHs with masses as low as $2 \times 10^3 \, \mathrm{M}_\odot$ ({\tt IC2-20-4}, {\tt IC2-100-4}) quickly sink towards the central MBH to form a tight binary. While only MBH-b (orange) and MBH-d (red) can end up in the central binary for low BH masses, every satellite MBH has a chance to form the central binary for satellite masses $\ge 2 \times 10^5 \, \mathrm{M}_\odot$ (indicated by the appearance of various colors in the solid lines).

All simulations with collisional (unsoftened) MBH-dark matter interactions show a continuous hardening of the semi-major axis due to MBH binary scouring. This effect is visible for the semi-major axes of the central binary (solid lines) as well as in the shrinking of the semi-major axis of the outer companions (colored surfaces). If the MBH-dark matter forces are softened, the forming binary MBHs do not harden and the semi-major axes are only reduced in three-body MBH interactions (i.e. MBH-b, orange, and MBH-d, red, in {\tt IC2-20-soft-6}), causing $a(t)$ to be approximately piece-wise constant.

\begin{figure*}
	\includegraphics[width=2.0\columnwidth]{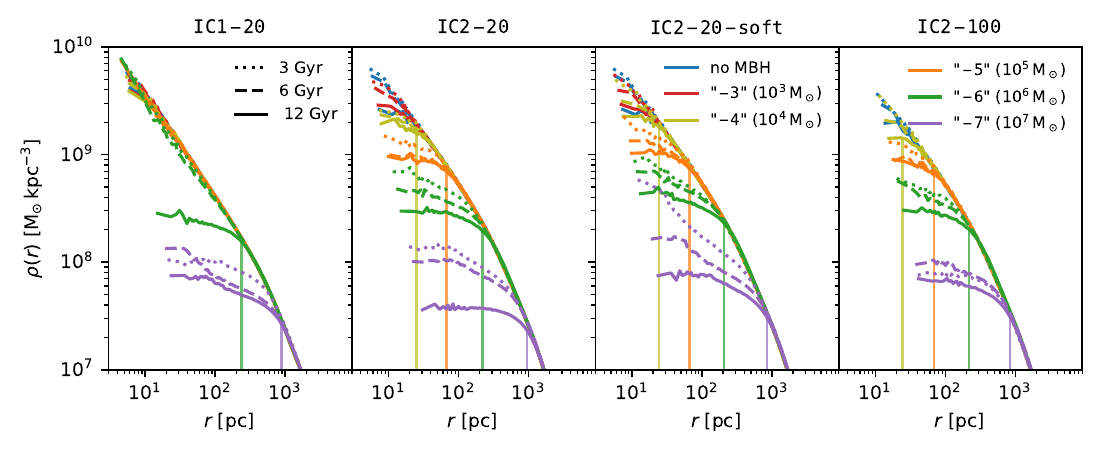}
    \caption{The dark matter densities profiles for the simulations presented in Fig. \ref{fig:sinking_dm} and \ref{fig:semimajor_axis} after 3 Gyr (dotted), 6 Gyr (dashed) and 12 Gyr (solid) of evolution for central MBHs with $M_{\bullet, \rm c}=10^3$ (red),  $10^4$ (green), $10^5$ (orange), $10^6$ (blue), and $10^7$ (purple) $\mathrm{M}_\odot$. The comparison simulations for the respective initial conditions without MBHs are shown in blue. Simulations which have not formed a central MBH binary (e.g. {\tt IC1-20-2}, {\tt IC1-20-4}, {\tt IC1-20-5}, {\tt IC2-20-3}) show only a minor reduction in central density, similar to the no-MBH simulation. Systems forming central binaries and/or MBH ejection show clear central deviations (breaks)  from the no-MBH simulations for central MBHs $\gtrsim 10^5 \, \mathrm{M}_\odot$. {\tt IC-20-7} and {\tt IC2-200-7} have the lowest central densities (a reduction by more than two orders of magnitude) as their MBHs are the most massive and the host MBHs have been ejected in recoil events. On the x-axis, the radii $r_{10 M_{\bullet}}$ enclosing a total mass of 10 times the initial central MBH mass are indicated with vertical lines. Typically, inside these radii the density profiles break and become flatter than the no-MBH profiles.}
    \label{fig:density_dm_new}
\end{figure*}

\begin{figure}
	\includegraphics[width=1.0\columnwidth]{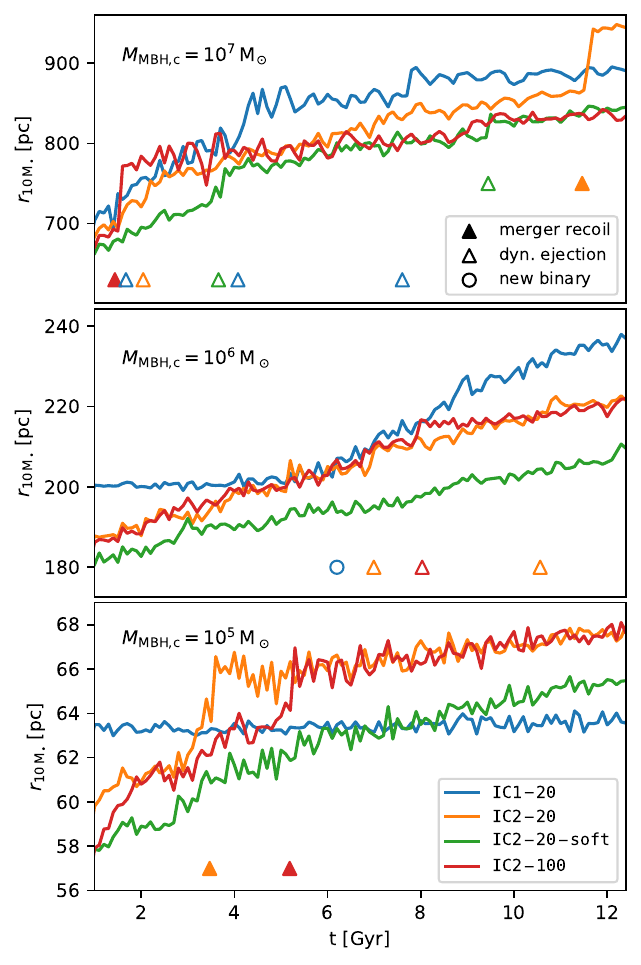}
    \caption{The time evolution of the radius enclosing dark matter with 10 times the central MBH mass $M_{\bullet, \rm c}$ for models with $M_{\bullet, \rm c}=10^7$ (top), $10^6$ (middle) and $10^5 \, \mathrm{M}_\odot$ (bottom). This radius is a good proxy for the break radius in the dark matter profile (see Fig. \ref{fig:density_dm_new}). Sudden increases in this radius are caused by dynamical ejection events (open triangles, e.g. {\tt IC2-20-6}, orange, middle panel). The strongest increases are associated with merger recoils (filled triangles, e.g. {\tt IC2-20-7}, orange, top panel). MBH binary scouring and the repeated kicks of satellite MBHs lead to a continuous increase of this radius (e.g. {\tt IC1-20-6}, blue, middle panel). Without a central MBH binary the radius remains constant (e.g. {\tt IC1-20-5}, blue, bottom panel). In the softened dark matter case, the effects are generally weaker.}
    \label{fig:lagrangian_rad_new}
\end{figure}

\begin{figure*}
	\includegraphics[width=1.8\columnwidth]{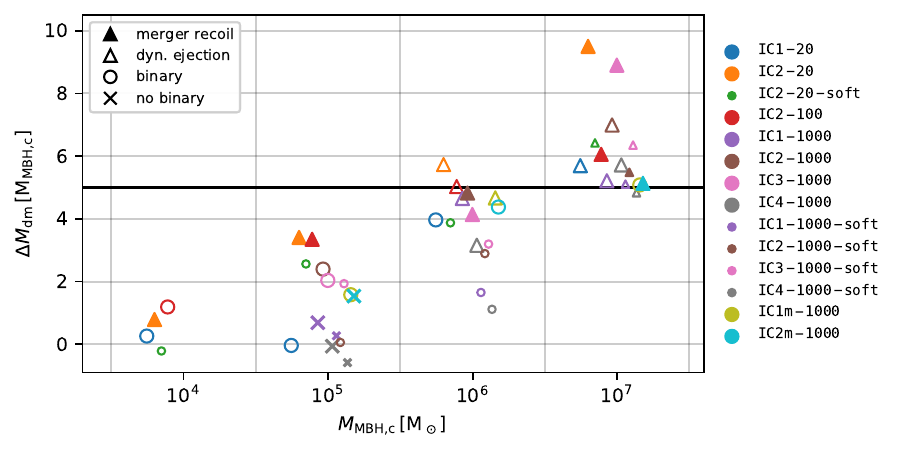}
    \caption{
    Missing dark matter mass ($\Delta M_{\rm dm}$) in units of the initial central MBH mass $M_{\bullet, \mathrm{c}}$ for four different initial MBH masses after 12 Gyr. The mass deficit $\Delta M_\mathrm{dm}$ is the mass difference within the respective final $r_\mathrm{10 M_\bullet}$ (see Figs. \ref{fig:density_dm_new} and \ref{fig:lagrangian_rad_new}) to the no-MBH simulations. 
    Here we also show additional realisations as listed in table \ref{tab:simulations} that were not shown in the previous figures. Simulations which do not form a binary (crosses) or have low central MBH masses (i.e. $10^4 \mathrm{M}_\odot$) show no or only a small mass deficit. With increasing central MBH mass, $\Delta M_{\rm dm}$ increases as more MBHs (repeatedly) sink and deplete the centre. Simulations in which the merger remnant of a MBH binary is ejected from the entire system (filled triangles) typically have the highest mass deficit. Simulations with softened MBH interactions typically lead to smaller missing masses. For the most massive MBHs ($10^7 \mathrm{M}_\odot$), the mass deficit can become almost as high as ten times the initial MBH mass. The black horizontal line indicates the expectation based on the assumption that all black holes sink and merge with the central MBH. Data points are artificially offset in x direction for better visibility.}
    \label{fig:missing_mass_8gyr}
\end{figure*}

These three-body interactions regularly result in exchanges of MBHs in the central binary, as indicated by the change of color. The most extreme case is {\tt IC2-100-6} with three exchanges. In these cases, a third satellite MBH sinks to a sub-parsec semi-major axis by loosing energy in interactions with the dark matter particles. Then, the central binary can either kick the sinking MBH or the sinking MBH replaces the binary partner of the central BH and the former binary partner receives a strong kick. These three-body interactions can either push satellite MBHs on wide orbits without subsequent sinking (e.g. open triangles in {\tt IC2-20-6}, MBH-b, orange, and MBH-a, blue, see also Fig. \ref{fig:sinking_dm}) or result in repeated sinking and interactions with the central MBH binary (e.g. MBH-d, red, in {\tt IC2-20-6}). As discussed in the previous section, the satellite MBHs can also be dynamically ejected from the entire system (e.g. MBH-e, purple, and MBH-a, blue, in {\tt IC1-20-7}). These ejections, indicated by open triangles, typically have a strong effect on the semi-major axis of the inner binary. 

If $a$ is small and the eccentricity of the central MBH binary becomes very high, gravitational wave emission drives the system towards a merger (filled black triangles in Fig. \ref{fig:semimajor_axis}). As an example, we highlight the path to the merger of two MBHs in {\tt IC2-20-5} in Fig. \ref{fig:merger_ejection}. In this case, the satellite MBH-b and MBH-d rapidly form a binary with a semi-major axis shrinking to $\sim 1 \, \rm pc$. In an encounter with the five times more massive central MBH, MBH-b is replaced by the central MBH and receives a kick. We show a representation of the MBH orbits during this process in the bottom left of Fig. \ref{fig:semimajor_axis}. The newly formed binary (MBH-d and the central MBH, red) forms a hierarchical triple with MBH-b (orange) and the semi-major axis of the inner binary shrinks to $\sim 100$ mpc within the first $\sim$ 3 Gyr. In contrast to the previous plots, where the semi-major axis was always computed with respect to the most massive central MBH, we compute $a$ for the outer triple companion (MBH-b) with respect to the centre of mass of the inner binary (MBH-d + central MBH). This avoids the oscillations in the semi-major axis of (MBH-b) that are visible in Fig. \ref{sec:dm_semi}. The outer MBH-b has a few close encounters with the central binary kicking it back to a larger semi-major axis. These encounters perturb the eccentricity of the inner binary (middle left panel). After the triple is established ($\sim$ 2.5 Gyr), the eccentricity (middle panel) shows periodic oscillations for the inner MBH binary which are typical for van Zeipel-Lidov-Kozai (ZLK) oscillations \citep{vonZeipel1910,Lidov1962,Kozai1962,Ito2019}. In the right panels of Fig. \ref{fig:semimajor_axis}, we show the evolution of the semi-major axes and eccentricities at higher time resolution shortly before the merger. The ZLK oscillations drive the inner MBH binary to increasingly higher eccentricities (low 1-$e$) and MBH-d merges with the central MBH at $\sim 3.57$ Gyr. The gravitational wave emission and circularisation of the orbits right before the merger are captured by the PN terms in the code but the timescale is too short to be visualised here. The merger remnant (grey, bottom right panel) is then ejected by its gravitational recoil kick and the third MBH (MBH-b, orange) becomes unbound and also leaves the central region to a wide orbit in the halo (see {\tt IC-20-5} in Fig. \ref{fig:sinking_dm}). The bottom right panel shows the orbit evolution around the time of the MBH merger with the merger remnant (grey) and the MBH (orange) leaving the centre.

As we discussed before, the evolution of the multiple MBH sub-systems at the halo centres has a stochastic component and neither the exact time of the merger nor the identity of the merging lower mass satellite MBH can be predicted from higher or lower resolution simulations. Even a different random realisation of the same initial conditions will most likely not result in the same MBH merger event. To assess the stability of the merger event prediction, we have shifted the initial positions of the central MBH in the initial conditions {\tt IC3-1000-7-perturbed} by displacements between dx $= 10^{-4} \, \rm pc$ and $30\, \rm pc$. In Fig. \ref{fig:butterfly_a}, we show the evolution of the semi-major axes (top left) and the eccentricities (bottom left) of the forming central MBH binaries for 23 different displacements with different colors. Independent of this displacement, the simulations show very similar hardening of the central MBH binary with similar final semi-major axes of about $\sim 200$ mpc (top right). However, in three of the realisations, the central MBH binary has merged without any clear connection to the initial displacement (filled triangles). The eccentricity evolution reflects this stochastic behaviour. The three simulations with the highest eccentricities result in MBH mergers. There is no correlation of initial displacement with final binary eccentricity (bottom right panel) and most realisations do not result in a MBH merger within a Hubble time. Also the satellite MBH which forms a binary with the central MBH can change with each realisation. This experiment highlights the complication of predicting MBH merger time scales even for the most accurate and highest resolution simulations. 

\subsection{Dark matter density distributions}
\label{sec:dm_dens}

As discussed in the introduction, coalescing MBHs in the centres of merging galaxies can lead to the formation of `cores' in the central stellar density profiles. Core formation is mainly caused by the transfer of energy from the MBHs to stars by dynamical friction and the slingshot ejections of stars (`scouring') through encounters with the MBH binary \citep[e.g.][]{2001ApJ...563...34M,2005LRR.....8....8M,Rantala_2017,2018ApJ...864..113R,2021MNRAS.508.4610F}. Most studies have focused on systems with comparable MBH masses. In general, however, the merger partners have unequal masses. According to simulations presented by \citet{2006ApJ...648..976M}, subsequent coalescence events with less massive MBHs are expected to result in a central mass deficit of order $\Delta M \sim 0.5 \, N_{\rm merger} \, M_{\rm BH, final}$, where $ N_{\rm merger}$ is the number of merger events and $M_{\bullet\mathrm{, final}}$ is the final MBH mass under the assumption that all MBHs merge without a merger recoil. The simulations discussed in this section have no stars. The dark matter, however, is assumed to directly interact with the central MBHs in the same way as the stellar population in the above studies and we therefore expect central mass deficits and the formation of dark matter density cores. In contrast to the simple expectation for the mass deficit in \citet{2006ApJ...648..976M}, we have shown that the MBH dynamics is more complex. The MBHs do not sink and merge one after the other but can interact and be ejected by dynamical interactions or recoil kicks which also affect the central density distribution \citep[e.g.][]{2004ApJ...607L...9M,2008ApJ...678..780G}. As shown in the previous section, MBHs can also sink to the centre multiple times. Although low-mass MBHs do not sink to the galaxy centre efficiently, many MBHs (or all, for the most massive MBHs) sink and form binaries or triples. This process is expected to create density cores even though real mergers of MBHs are the exception in the simulations. 

In Fig \ref{fig:density_dm_new}, we show the density profiles of the simulations presented in Figs. \ref{fig:sinking_dm} and \ref{fig:semimajor_axis} after 3, 6, and 12 Gyr of evolution. In addition, we show the evolution of the same initial conditions without MBHs in the dark matter halo centres (blue). Without MBHs, the dark matter density profiles evolve very little and only a small core region of order $\sim 20$ pc forms. This is a few times the gravitational softening length and is therefore expected. The density profile of simulations with $10^3$ and $10^4 \, \mathrm{M}_\odot$ central MBH mass (red and green) evolve very similarly and no density core forms. For higher MBH masses, core formation becomes visible if the MBHs can sink and form a binary at the centre (e.g. orange lines for {\tt IC2}). If no MBH binary is formed, the density profiles still do not differ from the case without MBHs. This is the case for simulations {\tt IC1-20} with central MBH masses $M_{\bullet, \rm c}\lesssim 10^5 M_\odot$ where no MBHs sink to the halo centre and form binaries. 

In all other cases, the central dark matter densities are reduced and flat-density cores form at the end of the simulations. The most striking example is {\tt IC2-20-7} (purple curve), where the central density drops by $\sim2$ orders of magnitude and leads to a constant density core with a size of almost $\sim 1 \, \rm kpc$. In this simulation, a long phase of a central binary evolution ($\sim12 \, \rm Gyr$) with exchanges, kicks and sinking is followed by a MBH merger with remnant ejection. Hence, all three mechanisms (binary formation, triple interactions with repeated kicks and sinking, and gravitational recoil) are at work here. If the binary MBH phase is short and the MBHs merge rapidly (e.g. {\tt IC2-100-7}), the core size and central density reduction is smaller. If the MBHs sink, even simulations with lower MBH masses (e.g. {\tt IC2-20-5} and {\tt IC2-100-5}) can lead to dark matter cores extending to $\sim 70 \, \rm pc$.

A dark matter core also forms if the dark matter - MBH forces are softened as in {\tt IC2-20-soft-7} but it is less pronounced than in the unsoftened case. Because the assumed force softening is very small ($\epsilon$ = 4 pc) dynamical friction and MBH binary formation at work also here, resulting in qualitatively similar behaviour. However, he strength of the effect is underestimated, MBH binary hardening and merging is suppressed, and the MBHs do not merge.

For the simple expectation that all MBHs sink to the centre, the missing mass estimated following \citet{2006ApJ...648..976M} would be proportional to the number of sinking (merger) events and the final black hole mass. Even though this simplified picture does not match the more complex behaviour in the simulations here, we empirically find that the radius enclosing ten times the initial central MBH mass, $M_{\rm dm}(r_{10 M_{\bullet}}) = 10 \times M_{\bullet, c}$, is a good indicator of the scale where the density profiles break and become shallower than in simulations without MBHs (as indicated by vertical lines in Fig. \ref{fig:density_dm_new}).

To get a better understanding of the time-evolution of the dark matter distribution, we show $r_{10 M_{\bullet}}$ as a function of time in Fig. \ref{fig:lagrangian_rad_new}. Significant, sudden increases can be attributed to merger induced kicks (filled triangles) or dynamical ejection (open triangles) that are strong enough to eject a MBH from the centre ($r > 5 \, \rm kpc$). Steady growth of $r_{10 M_{\bullet}}$ occurs if a binary is formed (e.g. {\tt IC1-20} in the $10^6 \, \mathrm{M}_\odot$ case, middle panel, blue line). However, the effect of idealised MBH sinking and merging cannot be separated from the effect of repeated weak kicks to radii comparable to $r_{10 M_{\bullet}}$ and subsequent sinking events here, since both typically occur simultaneously in our simulations. For example, simulation {\tt IC1-20-6} has a strong change in $r_{10 M_{\bullet}}$ after a binary forms at $\sim 6 \, \rm Gyr$ and $\sim 8$ sinking events from radii larger than $r_{10 M_{\bullet}}$ have happened, despite not having any mergers or dynamical ejections. In simulations with softened BH-dark matter interactions, the growth rate of $r_{10 M_{\bullet}}$ is generally smaller than in the unsoftened scenario. In agreement with Fig. \ref{fig:density_dm_new}, $r_{10 M_{\bullet}}$ does not increase if no binary is formed. 

It is difficult to single out the main driver of core formation here, since multiple processes happen simultaneously. However, from our simulations it becomes clear that the most significant effect usually comes from the ejection of a BH merger remnant, which can lead to a mass deficit within $r_{10 M_{\bullet}}$ of $\sim 1 - 4 \, M_{\bullet \rm ,c}$. Distinguishing between core scouring and the effect of repeated BH kicks is less clear. Even the comparison with the softened case, that cannot have core scouring below the softening length is insufficient here, because of the slightly longer sinking timescale compared to the unsoftened case. As a consequence, the heating through repeated kicks of the BHs is expected to be more inefficient. Hence, it is unclear if the slower growth of the core  radii can be attributed to the lack of core scouring or the fewer number of close encounters with other BHs as a consequence of less efficient dynamical friction.

We summarize our findings in Fig. \ref{fig:missing_mass_8gyr} where we show the mass difference at the Lagrangian radius $r_{10 M_{\bullet}}$ between the simulations with MBHs and simulations without MBH. In particular, we define the missing mass $\Delta M_{\rm dm}$ as
\begin{equation}
    \Delta M_{\rm dm} = M_{\rm dm}^{\rm no \, MBHs}(r_{10 M_{\bullet}}) - M_{\rm dm}^{\rm with \, MBHs}(r_{10 M_{\bullet}}),
\end{equation}

where the radius $r_{10 M_{\bullet}}$ is always computed for the simulations with MBHs such that $M_{\rm dm}^{\rm with \, MBH}(r_{10 M_{\bullet}}) = 10 \times M_{\bullet, c}$. Hence, the plot quantifies the amount of missing mass at the radius $r_{10 M_{\bullet}}$, that approximates the radius where density profiles with MBHs deviate from the simulations without MBHs (as shown in Fig. \ref{fig:density_dm_new}). To increase the sample, we are including the lower resolution simulations as well as the higher halo mass simulations with ten times higher halo mass as introduced in Tab. \ref{tab:simulations}. 

In agreement with the previous plots, we find that the mass difference scales with the masses of the involved MBHs. Simulations that had mergers usually lead to larger mass deficits, although an early merger, which leads to the ejection of the remnant, can stop the core formation process (e.g. {\tt IC2-100-7}, red filled triangle). The naive expectation based on \cite{2006ApJ...648..976M} would be a mass deficit of $\Delta M = 0.5 \times N_{\rm sink} \times M_{\rm MBH, final} = 5 \times M_{\bullet, c}$ (solid black line), under the assumption that all five satellite MBHs sink and merge without recoil. If the MBH mass increases, the number of MBH sinking events and dynamical encounters of satellite MBHs with a central binary become more frequent, which explains the relative increase of the missing mass. As expected, simulations without a binary do not have a measurable mass deficit. Hence, for high MBH masses, our simulations exceed the expectation according to \cite{2006ApJ...648..976M} due to the higher number of sinking events.

\begin{figure*}
	\includegraphics[width=1.8\columnwidth]{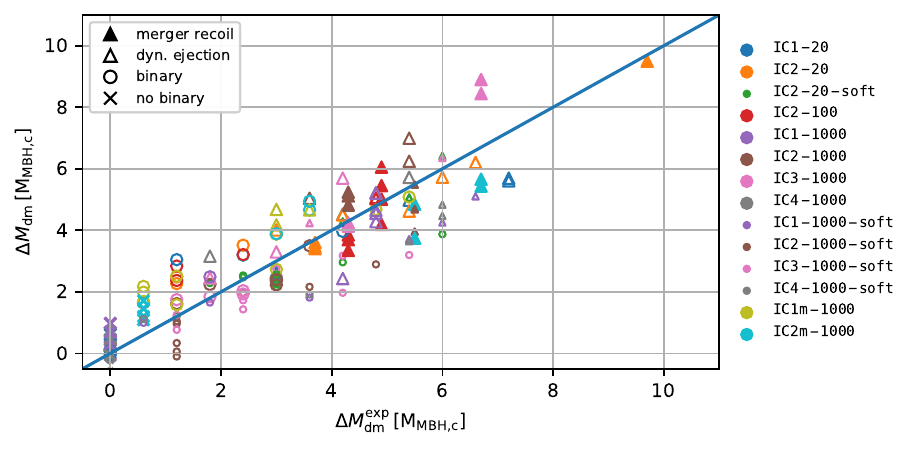}
    \caption{Comparison of the dark matter mass deficit inside $r_{10 M_{\bullet}}$ from the simulations, $M_{\mathrm{dm}}$ to the expectation according to Eq. \ref{eq:equation_exp}, $M_{\mathrm{dm}}^{\mathrm{exp}}$. The prediction based on the number of repeated sinking events and recoil kicks in the simulations reproduces the simulation results with a scatter of $\pm 2 M_{\bullet \, \rm c}$ and successfully explains the slope in Figure \ref{fig:missing_mass_8gyr}.}
    \label{fig:expected_missing_mass}
\end{figure*}

\begin{figure*}
	\includegraphics[width=2.0\columnwidth]{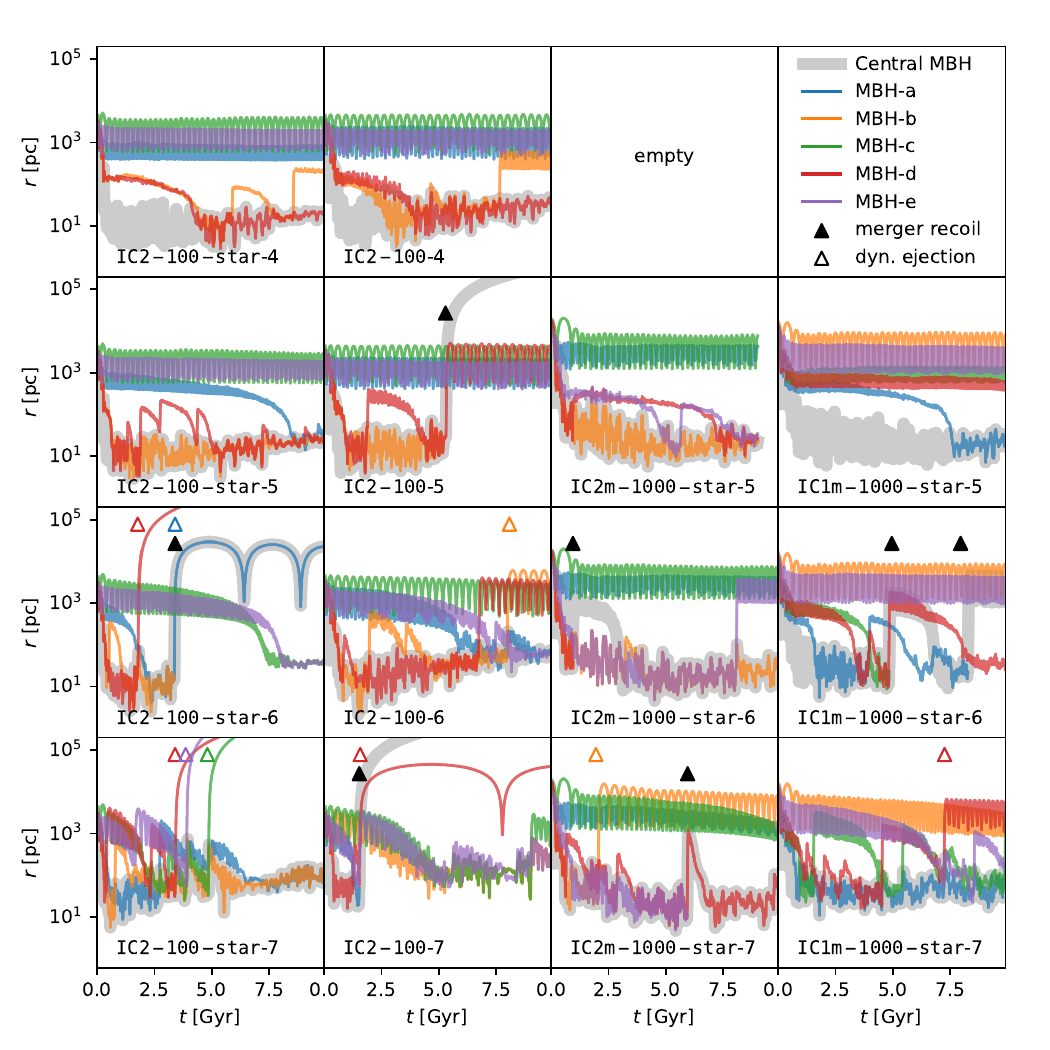}
    \caption{Radial distances of all MBHs (colored lines) from the (dark matter + stellar) density centre of the systems (similar to Fig. \ref{fig:sinking_dm}) as a function of time. We compare {\tt IC2} with stars (first comumn) to the already discussed case without stars (second column). In the third and fourth panel, we show the results for ten times larger dark matter halo masses. From top to bottom, the central MBH mass increases by factors of 10 from $10^4 \, \mathrm{M}_\odot$ to $10^7 \, \mathrm{M}_\odot$ (indicated by the grey band). In simulations with stars, MBHs sink much faster, especially in the central region that is dominated by stars. In the high mass halos ({\tt IC1m, IC2m}), kicked MBH merger remnants (black triangles) can not escape from the host halo anymore and sink back to the centre after the merger (e.g. {\tt IC1m-1000-star-6)}.}
    \label{fig:sinking_stars}
\end{figure*}

\begin{figure*}
	\includegraphics[width=2.0\columnwidth]{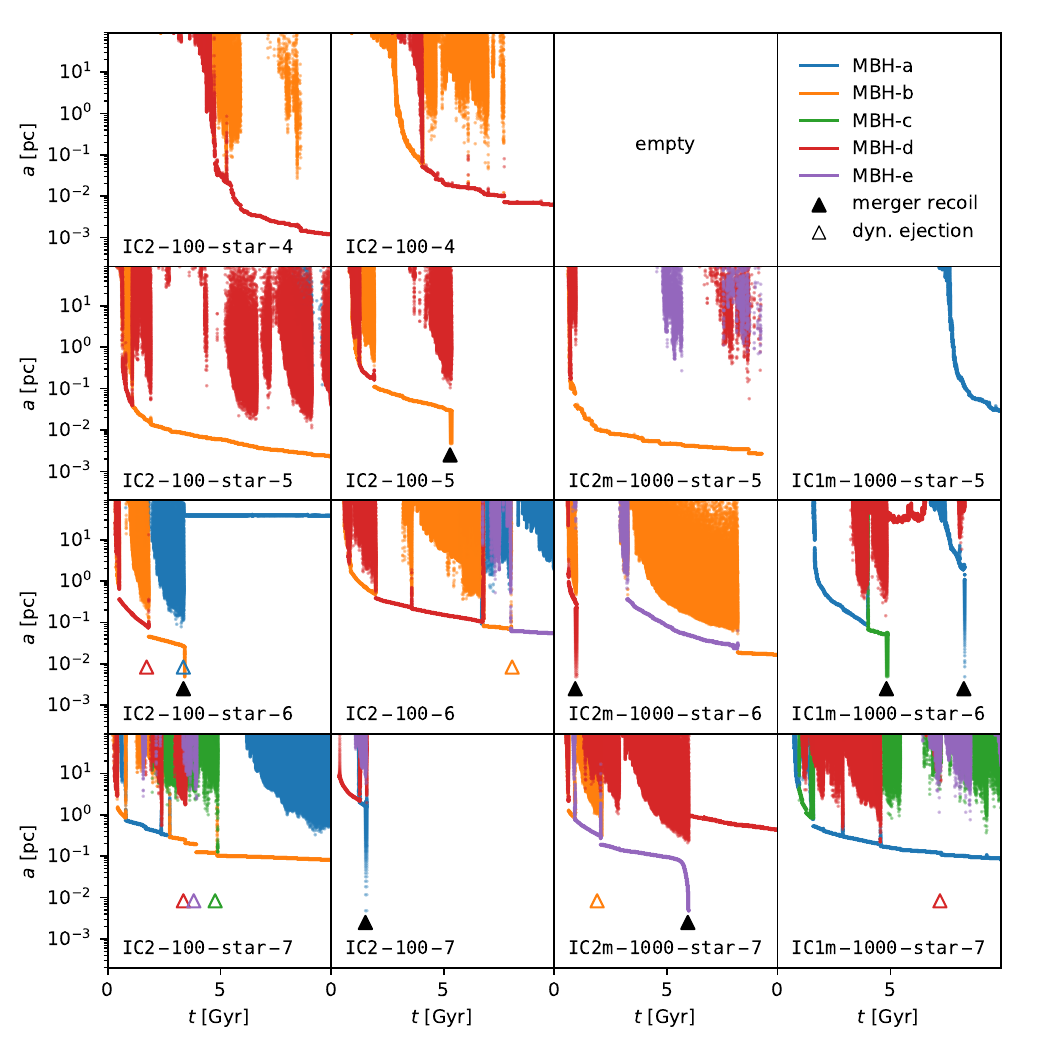}
    \caption{Semi-major axes, $a$, of the satellite MBHs and the more massive host MBHs as a function of time (similar to Fig. \ref{fig:semimajor_axis}, same plot without stars). Panels are arranged as in Fig. \ref{fig:sinking_stars}. Stars accelerate the hardening of the central binary and the semi-major axis is smaller than in the same simulations without stars. In simulations with larger halo masses (the third and fourth columns), the central MBH can undergo multiple mergers ({\tt IC1m-1000-star-6}) because it is not fully ejected from the host halo after a merger and can sink back to the centre.}
    \label{fig:semimajor_ax_stars}
\end{figure*}

With our simulation results, we can empirically make a more accurate estimate of the missing mass based on the actual number of sinking events and recoil kicks.  
For independent sinking events, we would expect a mass deficit of $\Delta M = 0.5 \times N_{\rm sink} \times M_{\bullet \rm , final}$. In contrast to the experiments presented in \cite{2006ApJ...648..976M}, in our simulations the central MBHs does not naturally grow by sinking events, because central satellite MBHs are usually kicked out again before they merge. Hence we use $ M_{\bullet, \rm final} = M_{\bullet \rm ,c} + M_{\bullet \rm ,s} = 1.2 \times M_{\bullet \rm ,c}$ as an approximation for the final mass, assuming that all sinking event are independent. We count a sinking event if a MBH sinks towards the centre from a distance that is greater than the typical value for $r_{10 M_{\bullet}}$ at the given MBH mass scale. We chose this radius because it represents the length scale at which the MBHs start to have an impact on the dark matter density. In particular we assume a threshold of $r_{\rm sink} = 65, 220$ and $900 \, \rm pc$ for central MBH masses of $10^5, 10^6$ and $10^7 \, \rm M_\odot$, respectively (see Fig. \ref{fig:lagrangian_rad_new}). For recoil ejections of MBH merger remnants, we empirically find a typical change in the missing mass of $\sim 1-4 \,M_{\bullet, \rm c}$. For simplicity, we assume that each recoiling merger remnant leads to a mass deficit of $\sim 2.5 \,M_{\bullet \rm ,final}$ and neglect the effect of dynamical ejections of satellite MBHs. Together, sinking satellite MBHs and BH merger recoils yield an expected mass deficit of 

\begin{equation}
\label{eq:equation_exp}
    \Delta M_{\rm  dm}^{\rm exp} \sim (0.5 \times N_{\rm sink} + 2.5 \times N_{\rm merger \, recoil}) \times (M_{\bullet \rm ,c} + M_{\bullet \rm ,s}) .   
\end{equation}

In Fig. \ref{fig:expected_missing_mass}, we show the comparison of the measured mass deficit $\Delta M_{\rm  dm}$ to the expected missing mass $\Delta M_{\rm  dm}^{\rm exp}$. We have included all dark matter simulations listed in Tab. \ref{tab:simulations} and combine data from $3, 6, 9$ and $12 \, \rm Gyr$. We find a correlation between the missing mass and the number of black hole sinking and merger events with a scatter around the expected mass deficit $\Delta M_{\rm  dm}^{\rm exp}$ smaller than $\pm 2 M_{\bullet \, \rm c}$. In general, simulations with softening (small symbols and green symbols) fall below the expectation. Despite its simplicity our model for $\Delta M_{\rm expected}$ successfully explains the slope in Fig. \ref{fig:missing_mass_8gyr}. 

The agreement is not expected to be perfect, because already in the more idealized experiments by \cite{2006ApJ...648..976M} the relation becomes less reliable for large numbers of sinking events, although they find that the linear scaling $\Delta M \propto  N_{\rm sink}$ holds for a variety of different density profiles and initial conditions. Also we emphasise that there are ambiguities in the definition of the sinking events, especially because it is difficult to determine the center of a galaxy with a large core. Also other assumptions for the "sinking radius" $r_{\rm sink}$ would be possible. Nevertheless, our estimate in Eq. \ref{eq:equation_exp} qualitatively and quantitatively explains the mass deficit given the knowledge of the simulation. Unfortunately, it cannot be used as a prediction as it is not known beforehand how the MBH subsystems will evolve and similar mass deficits can originate from different dynamical processes.

\section{Massive black holes in merging dark matter halos with galaxies}
\label{sec:dmstar}
In this chapter, we repeat the experiments presented in the previous section with central stellar component added to the dark matter halos. With this set of simulations, we investigate the effect of a galaxy on the sinking and hardening timescale as well as the resulting stellar and dark matter profiles. We use the same initial conditions for dark matter and their orbits as in Sec. \ref{sec:dm} and include central galaxies with masses of $2 \times 10^8 \, \mathrm{M}_\odot$ for the central and $4 \times 10^7 \, \mathrm{M}_\odot$ for the satellite halos. As introduced in Sec. \ref{sec:ics}, all galaxies have half-mass radii of 1 kpc (for the fiducial halo masses and the ten times higher halo masses). As the galaxies have a higher central density and the simulations are computationally more expensive we only use two resolutions, 100 and 1000 $\, \mathrm{M}_\odot$, for stars and dark matter with central MBH masses $M_{\bullet, \mathrm{c}} \geq 10^4 \, \mathrm{M}_\odot$. The simulations including a stellar component are indicated by "{\tt -star}" added to the simulation name. All interactions between the MBHs and star particles are unsoftened.

\begin{figure*}
\includegraphics[width=0.86\columnwidth]{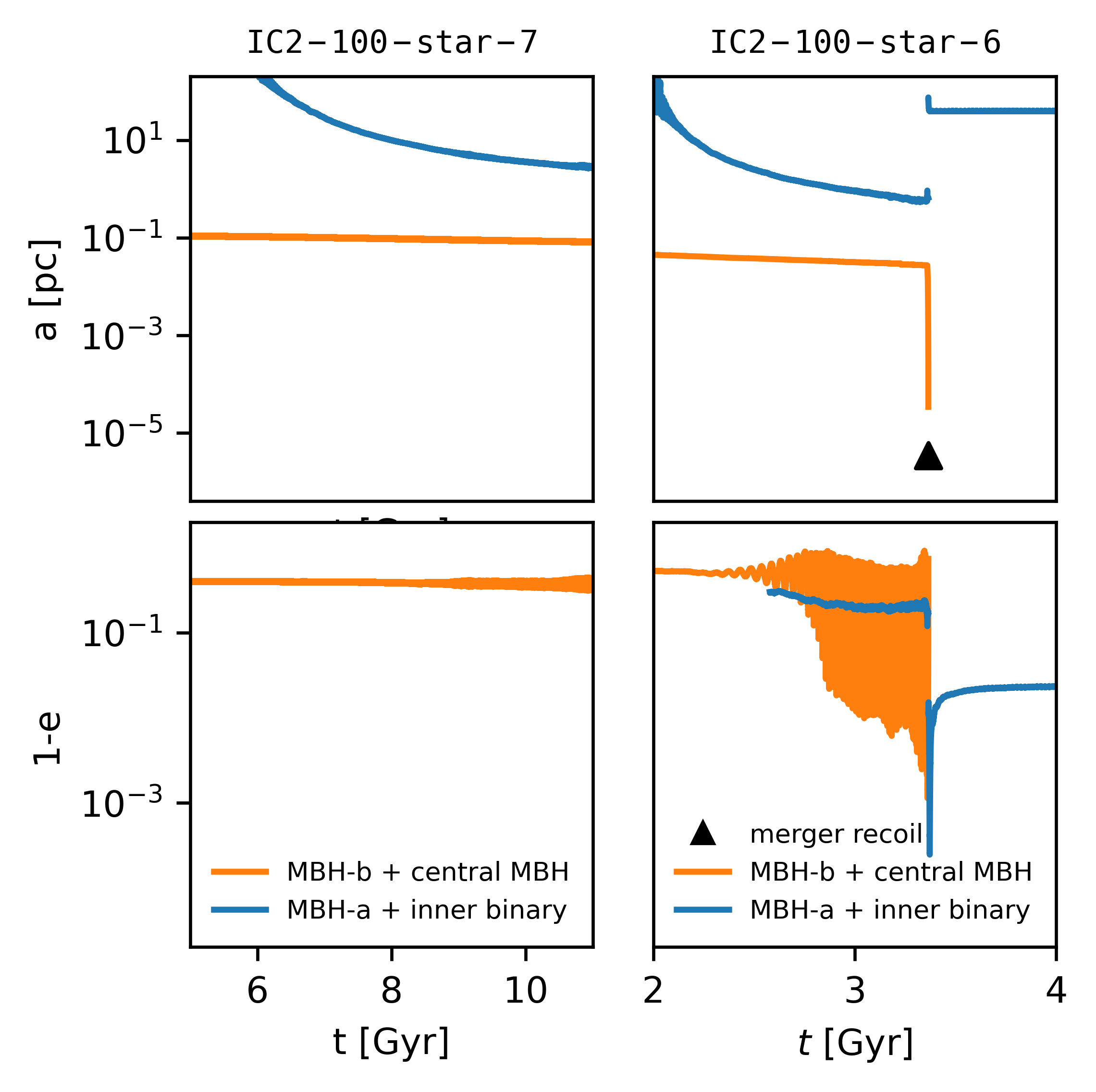}
	\includegraphics[width=1.14\columnwidth]{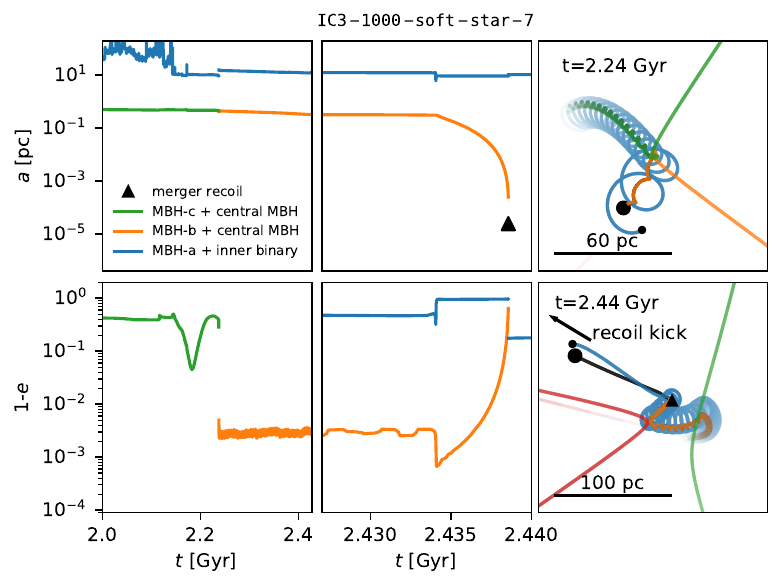}

    \caption{In {\tt IC2-100-star-7} (left), the inner binary has a low eccentricity of $e\sim 0.5$ and can not merge. In {\tt IC2-100-star-6} (middle), the outer triple companion (blue) in a long lived hierarchical triple excites periodic oscillations in $e$, eventually pushing the binary into the gravitational wave regime. The outer triple companion remains bound to the merger remnant. In the right panel, the exchange of the inner binary partner (green to initially unbound orange MBH, top right) increases $1-e$ by two orders of magnitude. The merger is then triggered by another close encounter with the red MBH (bottom right), pushing eccentricity above $e=0.999$.}
    \label{fig:ZLK_stars}
\end{figure*} 

\subsection{Massive black hole sinking}

In Fig. \ref{fig:sinking_stars}, we show the distance of the MBHs from the centre as a function of time, similar to Fig. \ref{fig:sinking_dm}. For better comparison, we show a set of simulations with stars (first column, {\tt IC2-100-star}) and the same orbital configuration with dark matter only (second column, {\tt IC2-100} is repeated from Fig. \ref{fig:sinking_dm}). The third and fourth columns show simulations with ten times higher halo masses, hosting galaxies with 1$\%$ of the halo mass.

In comparison to runs without stars, the satellite MBHs sink much faster, especially when they enter the inner  $r \lesssim 300 \rm \, pc$ that are dominated by the stellar component (e.g. MBH-a in {\tt IC2-100-star-6} in comparison to {\tt IC2-100-6}). However, satellite MBHs can remain on stable orbits in the halo outskirts if they spend most of their time outside this region (e.g. MBH-c and MBH-e in {\tt IC2-100-star-5}). The presence of stars also affects the sinking timescale of MBHs that have been kicked by a central binary, which return to the centre of the galaxy faster. Despite the shorter sinking timescales, the presence of stars does not always lead to more or faster mergers than in the simulations without stars ({\tt IC2-100-7} vs. {\tt IC2-100-star-7}).

Qualitatively, simulations with more massive halos behave in a similar way, although here the merger recoil kick velocity is not high enough to eject the remnant from the galaxy (grey lines and black triangles in {\tt IC2m-1000-star-6} and {\tt IC2m-1000-star-7}). As a consequence, MBHs can now merge with the central MBH multiple times, if the merger remnant sinks back to the centre. Because we are assuming zero MBH spin initially, this result might change if other MBH spins are used which can lead to higher recoil velocities \citep{2015PhRvD..92b4022Z}. We also do not find any dynamical ejections from the systems through three-body slingshots in the high mass case, likely because of the larger escape velocity of the more massive halos. As in the low halo mass case, MBHs can stay on wide orbits without sinking to the halo centre. Since the stellar-to-halo mass ratio is smaller in the high halo mass simulations and stars only dominate the inner $\lesssim 100 \, \rm pc$, the effect of the stars on the sinking timescales is less pronounced than in the low halo mass case.

An interesting case is {\tt IC2-100-star-6}, where the merger recoil in principle exceeds the escape velocity of the halo. However, the merger remnant drags along a third tightly bound black hole (MBH-a, blue) such that the resulting velocity of the binary (merger remnant and MBH-a) is not high enough anymore to escape from the halo. As a result, the MBH merger remnant orbits at $\sim 1 - 20 \rm kpc$, still bound to its former triple companion. We will discuss this case in more detail in the following section.

\subsection{Massive black hole binary formation, triple interactions, and mergers}

As the time-evolution of the semi-major axes in Fig. \ref{fig:semimajor_ax_stars} shows, MBH binaries harden significantly faster in the presence of a stellar component (e.g. MBH-d in {\tt IC2-100-star-6} vs. {\tt IC2-100-6}). Together with the shorter sinking timescale that causes kicked MBHs to return to the centre faster where they can scatter with the central MBH binary again, these effects lead to an enhanced probability for MBHs to merge. 
For example, among 13 simulations with a central MBH mass of $10^6 \, \mathrm{M}_{\rm \odot}$ and a stellar component, $\sim 60 \%$ lead to a merger (as opposed to $\sim 14 \%$ in similar simulations without stars). Except for simulation {\tt IC2-100-star-4}, where the binary forms late, the semi-major axis in the presence of stars is always smaller than in simulations with only dark matter. However, even in the presence of stars, the merger process is stochastic and crucially depends on eccentricity. For example, despite forming a MBH binary with $0.1 \rm \, pc$ separation, simulation {\tt IC1-100-star-1e7} does not lead to a merger because the orbit is not eccentric enough ($e \sim 0.5$, see left panel of Figure \ref{fig:ZLK_stars}). On the other hand, MBHs in simulation {\tt IC2-100-7} without stars can merge despite their initially large semi-major axis through a collision at high eccentricity. 

As a result of the higher escape velocity of the ten times more massive halos, the MBH merger remnants in the higher halo mass runs are not fully ejected from the host galaxy. Hence, satellite MBHs can remain bound to the merger remnant (e.g. MBH-d in {\tt IC2m-1000-star-7}) or rapidly form a new binary that in some cases even merges a second time ({\tt IC1m-1000-star-6}).

\begin{figure*}
	\includegraphics[width=2\columnwidth]{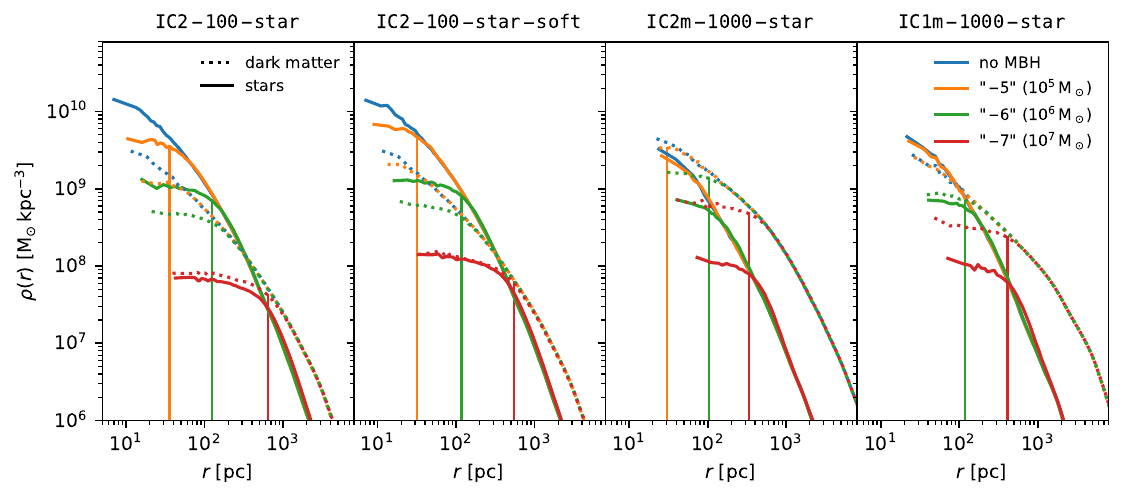}
    \caption{Comparison of the stellar and dark matter densities after $t = 9 \rm \, Gyr$. Similar to the dark matter simulations, the density profiles start to deviate from the simulations without BH at $r_{10 \, \rm M_\bullet}$ (indicated by vertical lines).  The impact of BH sinking and scouring is generally larger on the stellar component than on the dark matter profile, despite the unsoftened dark matter - MBH forces.}
    \label{fig:dmstar_density}
\end{figure*}

As discussed in section \ref{sec:dm_semi}, one way to excite eccentricity are ZLK oscillations in a hierachical triple. An example for this process is shown in the middle panel in \ref{fig:ZLK_stars}. Initially, the eccentricity of the inner binary is low and approximately constant at $e \sim 0.5$. While the outer triple companions semi-major axis shrinks, the eccentricity of the inner binary starts to oscillate and exceeds $e \sim 0.999$ (middle panel, around $t \sim 3.36 \rm \, Gyr$). This leads to an extremely small pericenter distance, pushing the binary into the gravitational wave driven regime where the binary merges rapidly. In this case, the BH merger recoil does not cause an ejection of the merger remnant from the host halo, even though it exceeds the escape velocity. Instead, the merger remnant (MBH-b + central MBH) remains bound to the former outer triple companion (MBH-a, blue line) that it shares the momentum of the recoil kick with. In addition, the semi-major axis of the resulting binary increases from $\sim 1 \, \rm pc$ to $\sim 30 \, \rm pc$, absorbing some of the kinetic energy of the recoil kick. This new binary is kicked to an orbit between $\sim \, 20 \rm \, kpc$ and $\sim 1 \, \rm \, kpc$, where it does not sink back to the galactic centre within the Hubble time (see Fig \ref{fig:sinking_stars}). Because the other MBHs have either not sunken to the centre yet (MBH-c and MBH-e) or have been ejected by dynamical interactions (MBH-d), there is no BH in the centre of the halo for $\sim 5 \rm Gyr$.

Another path to high eccentricities are scattering events with unbound BHs. As shown in the third column in Fig. \ref{fig:ZLK_stars}, the initially unbound MBH-b (orange) replaces MBH-b (green), that is initially in a tight binary with the central MBH. As a result, $1-e$ increases by approximately two orders of magnitude. Scattering events like this can in general also change the semi-major axis. A close encounter with another MBH (red, see bottom right panel for the trajectory and the fourth column for $a$ and $e$) finally excites enough eccentricity to trigger the merger. The eccentric orbit of the binary circularises through gravitational wave emission and the semi-major axis drops to the merger criterion (at $10 \, r_{\rm s}$) within roughly ten million years. In our simulations, all mergers are assisted by interactions among at least three BHs.

\subsection{Dark matter and stellar density distributions}

The density profiles of the simulations with stars and dark matter are shown in Fig. \ref{fig:dmstar_density}. Here, we also show the density profiles of a simulation with unsoftened star-BH forces but softened BH-dark matter interactions ({\tt IC2-100-star-soft}). Similar to the dark matter simulations (see Fig. \ref{fig:density_dm_new}), the density profiles in the presence of MBHs start to deviate from our comparison simulations at around $r_{10 \, \rm M_\bullet}$, where both the stellar and dark matter density profiles flatten. In simulations with stars, $r_{10 \, \rm M_\bullet}$ refers to the radius enclosing a total mass (stars and dark matter) of $10 \, \rm M_{\bullet, \rm c}$. 

The simulations with low halo mass, ({\tt IC2-100-star} and {\tt IC2-100-star-soft}), are initially dominated by stars inside a radius of $\sim 300 \, \rm pc$. Because the central stellar density is higher, stars are initially more susceptible to dynamical effects from MBH coalescence and their density drops significantly. Even though the break radius is similar for stars and dark matter, stars seem to be in general more affected and the central stellar density can drop below the dark matter density as in {\tt IC2-100-star-7}. At the end of the simulation this galaxy is now dark matter dominated at all radii. The effect is even more pronounced in the higher halo mass simulations, where the stellar density drops below the central density by a factor of $\sim 5$ for the most massive MBHs. Nevertheless, also in the high halo mass case the break radii for stellar and dark matter components are similar.

\begin{figure*}
	\includegraphics[width=2\columnwidth]{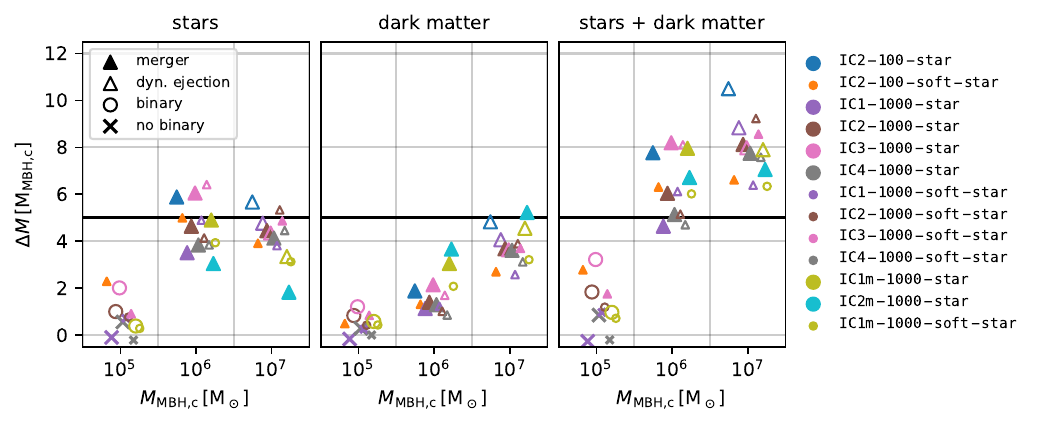}
    \caption{Missing mass inside $r_{10 \rm M_\bullet}$ as a function of central MBH mass, separated into stars and dark matter at $t = 10 \, \rm Gyr$. For $M_{\bullet, \rm c}<10^6 \, \rm M_\odot$, the missing stellar mass is larger than the missing dark matter mass. For the most massive MBHs, the effect of the scouring extends beyond the radii dominated by the stellar component and stars and dark matter have similar contributions to $\Delta M$. }
    \label{fig:missingmass_stars_tmp}
\end{figure*}

The fact that the stellar and dark matter distribution are affected differently is also visible in Fig. \ref{fig:missingmass_stars_tmp}, where we show the missing stellar (left), dark matter (middle) and total mass inside $r_{10 \, \rm M_\bullet}$. For MBH masses $M_{\bullet, \rm c} \lesssim 10^6 \rm \, M_{\odot}$, most of the missing mass is contributed by stars. This is not surprising, because the stellar component dominates the central region and is hence more susceptible to BH scouring. This is not the case anymore for the most massive BHs, where $r_{10 \, \rm M_\bullet}$ can extend into the regime where the galaxy is dark matter dominated. Hence, once the stars are removed by the MBHs, also dark matter can be scoured from the centre of the galaxy converging to comparable central mass deficits for dark matter and stars. 

Similar to the dark matter only case (see Fig. \ref{fig:missing_mass_8gyr}), we find a strong correlation between total missing mass $\Delta M$ and MBH mass $M_{\bullet, \rm c}$. Because there are typically more sinking events and mergers in the presence of stars, the total missing mass is typically larger than in the simulations without stars. The typically larger number of sinking events also makes the relative impact of mergers weaker and the largest mass deficits are not necessarily related to a merger ejection (e.g. {\tt IC2-100-star-7} had no mergers but three dynamical ejections and $\sim 11$ sinking events).

\begin{figure*}
	\includegraphics[width=2\columnwidth]{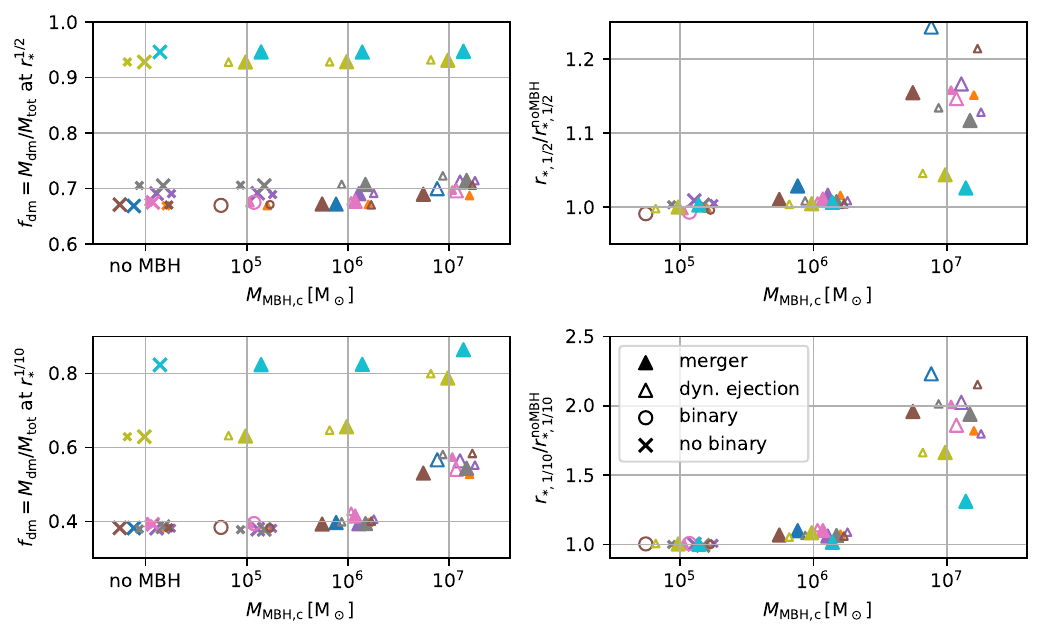}
    \caption{Shown is the dark matter fraction $f_{\rm dm}$ inside the half-light radius (top left) and the radius enclosing $10\%$ of the stellar mass (bottom left) at time ($t = 9 \, \rm Gyr$). Massive BHs have can change the central baryon fraction significantly, but also lead to per cent level changes on the scale of the half-light radius. The change in stellar density also changes the half-light radius (top right) as well as the Lagrangian radius enclosing $10\%$ of the stellar mass (bottom right). The colors of the symbols correspond to the legend in Fig. \ref{fig:missingmass_stars_tmp}. }
    \label{fig:stars_environment}
\end{figure*}

As shown in Fig. \ref{fig:stars_environment}, the efficient removal of stellar mass from the centre also leads to an increase in dark matter fraction $f_{\rm dm}$. While MBHs with $\lesssim 10^6 \rm \, M_{\odot}$ have not enough impact on the density distribution on large scales, the most massive MBH can change the dark matter fraction within the stellar half-light radius by a few per cent. The effect is more pronounced in the centre of the galaxy (i.e. the radius that encloses 10~per~cent of the stellar mass), where the dark matter fraction can increase from $\sim 40$ to $\sim 60$ per cent. At the same time, the half-light radius changes, because material is redistributed from the centre to the outskirts, as shown in the right panels of Figure \ref{fig:stars_environment}. Also here the effect is most visible in the central region, but can also change the galaxy half-light radius by $\sim 20$ per cent.

\subsection{Do merger remnants carry dark matter and stars?}

When MBHs are ejected due to merger recoil or dynamical interactions with the central binary, stars or dark matter can in principle remain bound. However, in our simulations, we only find one case where substantial mass remains bound to a merger remnant. In simulation {\tt IC2-100-7}, the central $10^7 \, \mathrm{M}_\odot$ MBH merges through a collision at extremely high eccentricity, before scouring can eject mass from the vicinity of the binary (see Fig. \ref{fig:semimajor_ax_stars}). After the merger, the remnant is ejected from the halo and keeps a dark matter cluster of $\sim 4300 \, \mathrm{M}_\odot$. Because the initial conditions for this simulation are without stars (see chapter \ref{sec:dm}), only dark matter can remain bound the the recoiling BH here. Due to the significantly reduced densities as a results of core scouring, we consider it unlikely - for conditions similar to our simulation setup - that typical MBH merger remnants carry significant amounts of dark matter and stars when they are ejected from the host galaxy. This might be different for MBHs embedded in nuclear star clusters, which are not considered in this study.

\section{Discussion}
\label{sec:discussion}
Our simulations start from idealized initial conditions, but we find a variety of different phenomena that we expect to be important also in more realistic environments.

Most idealised and cosmological galaxy evolution simulations typically do not resolve dynamical friction on MBHs and rely on approximate models. Here, we test the impact of resolved, unsoftened interactions between MBHs and their dark matter and stellar environment. We find that these collisional interactions between BHs and dark matter accelerate the sinking and hardening of BH binaries. With this approach, we take into account the point-like nature of BHs, allowing for accurately computed close encounters of our dark matter particles (representing the dark matter phase space) and the BH with potentially large accelerations. Similar to the modelling of stars in typical galaxy simulations where stellar population particles are not representing individual stars but rather trace their (on large scales collision-less) phase space, our dark matter particles trace the dark matter field without challenging the assumption that dark matter is collision-less on large scales. In fact, the small scale interactions considered here are not in conflict with traditional approaches, as e.g. the derivation of the \cite{1943ApJ....97..255C} dynamical friction formula explicitly considers collisional deflections of dark matter particles by the point-like BH.

Resolving the cumulative effect of these encounters requires a high mass ratio between the MBHs and dark matter and star particles. With our simulations, we are confident that BH-dark matter/star scouring is well resolved down to MBH masses of $10^4 \, \mathrm{M}_\odot$ for our highest resolution of $20 \, \mathrm{M}_\odot$, where we find very good agreement in the hardening rates when comparing different resolutions (e.g. {\tt IC2-20-5} vs {\tt IC2-100-5}). At lower MBH to star/dark matter mass ratios, we start to see the effect of individual encounters in the evolution of the semi-major axis (e.g. the evolution of $a$ in {\tt IC2-100-4}, Fig. \ref{fig:semimajor_axis} is not completely smooth anymore, although the hardening rate is still similar to the higher resolution simulation {\tt IC2-20-4}, see also \cite{10.1093/mnras/259.1.115}). However, the sinking of MBHs is still well resolved, even for the smallest tested MBH masses. 

With the careful modelling of MBHs in their environment as presented here, we emphasize that the dynamics of MBH systems is difficult to predict. Even small perturbations of the initial conditions can make the difference between a system that merges quickly or forms a long-lived low eccentricity binary. This is a known problem for MBH binaries \citep{2020MNRAS.497..739N,2023arXiv230708756R}. We note here that our highest resolution simulations resolve the stellar components already at 100 $\rm M_\odot$ which is close to the natural resolution limits of individual stars. The presence of multiple MBHs, that can scatter and form subsystems, makes the dynamics even more complicated. This is a problem for many cosmological simulations, that typically trigger mergers already at large distances, where the fate of the BHs is still undecided. 

In agreement with studies that use approximate dynamical friction prescriptions \citep[e.g.][]{Pfister_2019, Ma_2021}, we not only find that it can be difficult for low mass MBHs to migrate to the centres of galaxies, but even if they sink it is difficult for them to remain at the center and merge  in a Hubble time. Low hardening rates for binaries and dynamical kicks in multiple MBH systems easily kick them out of the center, repeatedly. Even the central binary does not necessarily merge and can instead remain stable at sub-parsec semi-major axis. In our simulations most of the sinking MBHs do not do not contribute to the growth of the central MBH growth. Our results are in qualitative agreement with the cosmologically motivated semi-analytical study by \cite{Volonteri_2005}. 

Another problem for MBH growth through mergers is that merger remnants are easily ejected from the host \citep{Haiman_2004}. While remnants are almost always ejected from the host halo for the low mass halo, the merger remnant can sink back to the centre in the more massive scenario. However, we likely underestimate the recoil velocity because we assume zero MBH spin initially and do not follow the spin evolution through gas accretion. While these recoil kicks can in principle have velocities up to $\sim 5000 \, \rm km/s$, the expected velocities for typical spin and mass ratios are in the order of $\lesssim 500 \, \rm km/s$ \citep{2012PhRvD..85h4015L}. Hence, even higher halo masses can in general be affected .

While we see a clear trend with mass in the sinking time-scale, it is not so clear which MBH masses have the highest probability for a merger. In the low-mass MBH case, only a small amount of energy has to be removed from the binary to trigger a merger, but dynamical friction timescales are longer and the gravitational wave driven in-spiral requires smaller BH binary separations. On the other hand, for very massive MBHs, a binary can deplete the central region quickly such that scouring can become inefficient before the binary is hard enough to merge. Mergers of MBH binaries are typically assisted by interactions with additional MBHs that can excite eccentricity and remove energy. In our simulations, most mergers happen for central MBH masses of $M_{\bullet, \rm c} \sim 10^6 \, \mathrm{M}_\odot$. For the tested halos, this seems to be the optimal mass scale for both competing effects.

Our simulations clearly show that MBHs can change the density structure of the host galaxy on a scale that is proportional to their mass. Our experiments generalize the results presented in \cite{2006ApJ...648..976M}, where subsequent mergers of BHs in a stellar environment were studied. We find that MBHs, depending on their mass, can lead to stellar and dark matter cores of $\sim 70 \, \rm pc$ ($10^5 \, \mathrm{M}_\odot$) to $\sim \rm kpc$ ($10^7 \, \mathrm{M}_\odot$) size. The core radius is well correlated with the radius $r_{10 \, \rm M_\bullet}$ enclosing a mass of $10 \times M_{\bullet, \rm c}$. The strength of the effect depends on the number of mergers and dynamical MBH ejections and is more pronounced in the presence of repeated sinking events. Because massive MBHs are more likely to merge and repeatedly sink, we exceed the expectation based on the simplified experiments in \cite{2006ApJ...648..976M}. On the other hand, for low MBH masses ($M_{\bullet \rm, c} \lesssim 10^5 \, \rm M_\odot$), our mass deficits are smaller because MBHs usually do not sink to the centre. In agreement with e.g. \cite{2021MNRAS.502.4794N}, dynamical ejections of merger remnants lead to an additional mass deficit. The impact of BH dynamics on the dark matter profiles might be particularly interesting. As pointed out by \cite{Milosavljevic_2002}, a dark matter mass deficit might be long-lived, while stellar mass can be produced though star formation again, hiding the effect of a binary MBH in the early phase of galactic evolution. Based on our simulations, over-massive BHs might also be an additional path to dark matter cores in low mass galaxies that are inferred from observations \citep[e.g.][]{de_Blok_2010}.

Our simulations start from idealised initial conditions and are designed to lead to a quick merger within a few Gyrs. In a cosmological environment, the number, the timing and the galaxy and black hole mass ratios of mergers at high-redshift might be different. As pointed out by \citet{2018ApJ...864L..19T}, also the dark matter density structure of the host halo has an effect on the MBH dynamics such that the assumption of halos following Hernquist density profiles might be oversimplified. However, to sample a representative number of merger configurations it would be necessary to use a cosmological simulation with a large box size, which is not feasible at the moment. On the other hand, zoom simulations are a less controlled environment and also only sample one possible configuration at a time. As we have shown in the paper, varying the parameters of just one idealised set-up (in particular the BH mass, orbital configuration and the initial BH positions by a small perturbation) leads to a rich dynamics with a parameter space that is difficult to cover. Based on our controlled experiments presented here, the next step will be to examine some of the effects in cosmological zoom simulation.

Furthermore, including gas in simulations can lead to a clumpy structure of the interstellar medium. As pointed out by \cite{Ma_2023}, the sinking timescales in simulated, clumpy, high-redshift galaxies might be longer compared to the same galaxy without clumps and spherical symmetry. On the other hand, even though the impact of gas drag is generally expected to be small, it might be important at high-redshift \citep{Chen_2021} and accelerate the sinking process. However, processes like the effect of radiation on the dynamical friction wake can in principle reverse the effect of dynamical friction and accelerate the sinking object instead \citep{2017ApJ...838..103P}. Hence, it is in general important to consider effects beyond dynamical friction from dark matter and stars \citep[e.g.][]{Bortolas_2020, Tamburello_2016}. The presence of nuclear star clusters or a dense stellar component surrounding sinking BHs might also speed up the sinking process \citep{Pfister_2019, 2018MNRAS.475.4967T}. It is also important to note that high-redshift galaxies are likely not relaxed systems and do not have a well defined dynamical center, which makes sinking even harder. 

For the evolution of BH binaries, the gaseous circum-binary discs can have an important effect on the eccentricity evolution. For equal mass binaries, studies by \cite{D_Orazio_2021} find that the eccentricity of the binary either converges to $e \sim 0$ or $e \sim 0.4$, depending on initial eccentricity (similar results are reported by \citealp{Zrake_2021} and \citealp{Siwek_2023}). As we have discussed in the paper, eccentricity is crucial for triggering merger processes. Hence, resolving the physics of the circum-binary discs might have important impact on the evolution of the eccentricity, semi-major axis and the merger process. However, since our study extends beyond the binary regime and can involve multiple BHs, it is not clear how gas changes the dynamics. 

Compared to previous simulations in the literature, our simulations significantly improve the accuracy of dynamical interactions between MBHs and their stellar and dark matter environment, but neglect the potentially important impact of gas physics and employ idealized initial conditions. In future work we will study the interaction between MBHs and a structured and turbulent resolved multi-phase interstellar medium (ISM) as well as realistically clustered stellar populations with the \textsc{GRIFFING} ISM model presented in e.g. \citet{2020ApJ...891....2L} and \citet{2022MNRAS.509.5938H}.

\section{Summary and Conclusions}
\label{sec:conlusion}

We present an idealised  high resolution numerical study of the sinking and merging of MBHs with masses of $10^3 - 10^7 \, \mathrm{M}_\odot$ in multiple mergers (typical mass ratios of 5:1) of low mass dark matter halos and galaxies ($4\times 10^8 \, \mathrm{M}_\odot \lesssim \mathrm{M}_{\mathrm{halo}} \lesssim 2\times 10^{10} \, \mathrm{M}_\odot)$. The simulations are carried out with the \textsc{Ketju} code in a combination of the \textsc{Gadget} tree solver with accurate regularised integration around the MHBs. The highest mass resolution for dark matter particles is 20 $\mathrm{M}_\odot$  and 100 $\mathrm{M}_\odot$ for stellar particles. This is close to the fundamental limit of resolving individual stars. The interaction of dark matter particles and stars with the MBHs is unsoftened allowing for an accurate treatment of dynamical friction and scattering of dark matter/stars by MBH binaries or mutiples. The simulations include post-Newtonian correction up to order 3.5 for MBH interactions allowing for coalescence by gravitational wave emission and a prescription for gravitational recoil kicks. With this study we aim at a better understanding of the evolution of MBH populations representing various seeding scenarios in merger dominated low mass halo envrionment resembling conditions in the early Universe. Our main findings can be summarized as follows:

\begin{itemize}
\item Low mass MBHs ($ \lesssim 10^5 \, \mathrm{M}_\odot$) in general do not sink to the halo or galaxy centres efficiently. For special orbitals configurations even low mass MBH sinking is possible.

\item If MBHs sink to the halo centre, they can form binaries or triples. In the case of resolved (not force softened) dark matter-MBH interactions, the semi-major axes can harden through dark matter particle slingshot ejection. For softened dark matter - MBH forces, a hard binary can only efficiently lose energy through interactions with other MBHs or stars. 

\item Binary MBH mergers are usually triggered by a third MBH which excites a high eccentricity in the inner binary and pushes it into the gravitational wave driven regime. Mergers are rare and often require long periods of time (several 100 megayears to gigayears), but we find mergers at all MBH masses $> 10^4 \, \mathrm{M}_\odot$ with the highest probability for mergers at a MBH mass scale of $\sim 10^6 \, \mathrm{M}_\odot$.

\item Due to the stochastic nature of close \textit{N}-body encounters, it is difficult to predict the long-term evolution of a halo/galaxy with multiple MBHs. Even small changes to the initial conditions can make the difference between a long-lived, low-eccentricity binary and a rapid merger. 

\item MBH merger remnants are typically ejected by gravitational recoil kicks from low-mass host halos. For the higher halo mass ($2 \times 10^{10} \, \mathrm{M}_\odot$), the kick velocity for non-spinning MBHs is too low to eject the remnant from the halo. If a MBH binary is present, it is common that this central binary kicks other MBHs during strong three-body interactions to wide orbits or even out of the halo. Because a significant fraction of MBHs that sink to the centre do not merge and merger remnants have a high chance to get displaced or ejected from the halo centre, this poses an additional challenge to merger-assisted MBH seed growth.

\item If a merger happens in a hierarchical triple, the recoiling merger remnants can in principle remain bound to the outer triple companion. In this case, a binary BH is ejected from the galaxy. Stellar or dark matter mass usually does not remain bound to the recoiling merger remnant.

\item Sinking MBHs and BH binaries produce a core in the central stellar and dark matter density. Together with merger recoil ejections or dynamical ejections of satellite MBHs, this leads to a mass deficit of up to $\sim 10 \, M_{\bullet \rm ,c}$ inside the core radius that is well approximated by the radius $r_{10 \, \rm M_\bullet}$. Consistent with the more idealized findings of \cite{2006ApJ...648..976M}, we find that the mass deficit scales approximately linearly with the number of MBH sinking events. 

\item MBHs can lead to flat density cores of up to $\sim 1 \, \rm kpc$ and change the central dark matter fractions as well as the stellar half-light radius. In extreme cases, this effect turns a galaxy that is initially dominated by stars in the centre into a dark matter-dominated system.

\end{itemize}

\section*{Acknowledgements}

We thank Silvia Bonoli, Volker Springel and Marta Volonteri for valuable discussions and scientific input. T.N. acknowledges support from the Deutsche Forschungsgemeinschaft (DFG, German Research Foundation) under Germany’s Excellence Strategy - EXC-2094 - 390783311 from the DFG Cluster of Excel- lence "ORIGINS". M.M. and P.H.J. acknowledge the support by the European Research Council via ERC Consolidator Grant KETJU (no. 818930) and the support of the Academy of Finland grant 339127. Computations were performed on the HPC system Raven, Cobra and Freya at the Max Planck Computing and Data Facility. C.P. and T.N. acknowledge the computing time granted by the LRZ (Leibniz-Rechenzentrum) on SuperMUC-NG under project numbers pn72bu.

\section*{Data Availability}

 The data will be made available based on reasonable request to the corresponding author.



\bibliographystyle{mnras}
\bibliography{literature} 



\newpage
\appendix

\bsp	
\label{lastpage}
\end{document}